\documentclass[aps,prl,twocolumn,superscriptaddress]{revtex4-1}
\pdfoutput=1

\usepackage{amsfonts}
\usepackage{amssymb}
\usepackage{amsmath}
\usepackage{upgreek}
\usepackage{color}
\usepackage{lineno}
\usepackage{graphicx}
\usepackage[colorlinks=true,citecolor=blue,linkcolor=blue,urlcolor=blue]{hyperref}

\newcommand{\beginsupplement} {
    \setcounter{table}{0}
    \renewcommand{\thetable}{S\arabic{table}}
    \setcounter{figure}{0}
    \renewcommand{\thefigure}{S\arabic{figure}}
    \setcounter{equation}{0}
    \renewcommand{\theequation}{S\arabic{equation}}
}

\newcommand*\patchAmsMathEnvironmentForLineno[1]{
  \expandafter\let\csname old#1\expandafter\endcsname\csname #1\endcsname
  \expandafter\let\csname oldend#1\expandafter\endcsname\csname end#1\endcsname
  \renewenvironment{#1}
     {\linenomath\csname old#1\endcsname}
     {\csname oldend#1\endcsname\endlinenomath}} 
\newcommand*\patchBothAmsMathEnvironmentsForLineno[1]{
  \patchAmsMathEnvironmentForLineno{#1}
  \patchAmsMathEnvironmentForLineno{#1*}}
\AtBeginDocument{
\patchBothAmsMathEnvironmentsForLineno{equation}
\patchBothAmsMathEnvironmentsForLineno{align}
\patchBothAmsMathEnvironmentsForLineno{flalign}
\patchBothAmsMathEnvironmentsForLineno{alignat}
\patchBothAmsMathEnvironmentsForLineno{gather}
\patchBothAmsMathEnvironmentsForLineno{multline}
}

\newcommand{\mytitle}{Dynamical slowing down in an ultrafast photo-induced phase transition}

\newcommand{\MIT}{Massachusetts Institute of Technology, Department of Physics, Cambridge, Massachusetts 02139, USA.}
\newcommand{\Skoltech}{Skolkovo Institute of Science and Technology, Skolkovo Innovation Center, 3 Nobel Street, Moscow, 143026, Russia.}
\newcommand{\Geballe}{Geballe Laboratory for Advanced Materials, Stanford University, Stanford, California 94305, USA.}
\newcommand{\StanforAP}{Department of Applied Physics, Stanford University, Stanford, California 94305, USA.}
\newcommand{\StanfordMSE}{Department of Materials Science and Engineering, Stanford University, Stanford, California 94305, USA.}
\newcommand{\SIMES}{SIMES, SLAC National Accelerator Laboratory, Menlo Park, California 94025, USA.}
\newcommand{\APS}{Advanced Photon Source, Argonne National Laboratory, Argonne, Illinois, 60439, USA.}
\newcommand{\SLAC}{SLAC National Accelerator Laboratory, Menlo Park, California 94025, USA.}
\newcommand{\Linac}{Linac Coherent Light Source, SLAC National Accelerator Laboratory, Menlo Park, California 94025, USA.}
\newcommand{\DESY}{Center for Free-Electron Laser Science, DESY, Notkestra{\ss}e 85, 22607 Hamburg, Germany.}
\newcommand{\SYSU}{School of Electronics and Information Technology, Sun Yat-sen University, Guangzhou, Guangdong 510006, China.}
\newcommand{\Harvard}{Department of Physics, Harvard University, Cambridge, Massachusetts, 02138, USA.}

\begin{document}

\title{\mytitle}

\author{Alfred~Zong}
\affiliation{\MIT}
\author{Pavel~E.~Dolgirev}
\affiliation{\Skoltech}
\affiliation{\Harvard}
\author{Anshul~Kogar}
\affiliation{\MIT}
\author{Emre~Erge\c{c}en}
\affiliation{\MIT}
\author{Mehmet~B.~Yilmaz}
\affiliation{\MIT}
\author{Ya-Qing~Bie}
\thanks{Present address: \SYSU}
\affiliation{\MIT}
\author{Timm~Rohwer}
\thanks{Present address: \DESY}
\affiliation{\MIT}
\author{I-Cheng~Tung}
\affiliation{\APS}
\author{Joshua~Straquadine}
\affiliation{\Geballe}
\affiliation{\StanforAP}
\affiliation{\SIMES}
\author{Xirui~Wang}
\affiliation{\MIT}
\author{Yafang~Yang}
\affiliation{\MIT}
\author{Xiaozhe~Shen}
\affiliation{\SLAC}
\author{Renkai~Li}
\affiliation{\SLAC}
\author{Jie~Yang}
\affiliation{\SLAC}
\author{Suji~Park}
\affiliation{\SLAC}
\affiliation{\StanfordMSE}
\author{Matthias~C.~Hoffmann}
\affiliation{\Linac}
\author{Benjamin~K.~Ofori-Okai}
\affiliation{\SLAC}
\author{Michael~E.~Kozina}
\affiliation{\SLAC}
\author{Haidan~Wen}
\affiliation{\APS}
\author{Xijie~Wang}
\affiliation{\SLAC}
\author{Ian~R.~Fisher}
\affiliation{\Geballe}
\affiliation{\StanforAP}
\affiliation{\SIMES}
\author{Pablo~Jarillo-Herrero}
\affiliation{\MIT}
\author{Nuh~Gedik}
\email[Correspondence to: ]{gedik@mit.edu}
\affiliation{\MIT}

\date{January 22, 2019}

\begin{abstract}
Complex systems, which consist of a large number of interacting constituents, often exhibit universal behavior near a phase transition. A slowdown of certain dynamical observables is one such recurring feature found in a vast array of contexts. This phenomenon, known as critical slowing down, is well studied mostly in thermodynamic phase transitions. However, it is less understood in highly nonequilibrium settings, where the time it takes to traverse the phase boundary becomes comparable to the timescale of dynamical fluctuations. Using transient optical spectroscopy and femtosecond electron diffraction, we studied a photo-induced transition of a model charge-density-wave (CDW) compound, LaTe$_3$. We observed that it takes the longest time to suppress the order parameter at the threshold photoexcitation density, where the CDW transiently vanishes. This finding can be quantitatively captured by generalizing the time-dependent Landau theory to a system far from equilibrium. The experimental observation and theoretical understanding of dynamical slowing down may offer insight into other general principles behind nonequilibrium phase transitions in many-body systems.
\end{abstract}

\maketitle


In a second-order symmetry-breaking phase transition, the spatial extent of fluctuating regions diverges close to the critical temperature, $T_c$. Correspondingly, the relaxation time of these fluctuations tends to infinity, a phenomenon known as critical slowing down \cite{Collins1989,Goldenfeld1992}. The phenomenology of slowing dynamics near a critical point is much more general: it has been observed in first-order transitions \cite{Horie1987,Zhu2018}, glasses \cite{Souletie1985,Lasjaunias1994}, dynamical systems \cite{strogatz2018}, and even microbial communities \cite{Veraart2012}. Its common occurrence makes it a robust signature of phase transitions in a vast array of complex systems \cite{Scheffer2009}.

Close to equilibrium, critical slowing down has been well characterized in condensed matter systems. Theoretically, it is described by a dynamical critical exponent, whose value depends on the dynamic universality class \cite{Goldenfeld1992}. Experimentally, the evidence comes from a vanishing rate of change in the order parameter close to $T_c$, with early reports in refs.~\cite{Collins1973,Iizumi1986,Horie1987}. While these measurements probe the slowing dynamics in the time domain, it can be observed in the frequency domain as well. For example, inelastic neutron scattering has revealed a narrowing quasi-elastic peak along the energy axis as $T_c$ is approached, indicating a suppressed relaxation rate of critical fluctuations \cite{Cailleau1979,Toudic1986,Press1974}. Moreover, if there is a collective mode associated with the phase transition, the mode softening in the vicinity of $T_c$ is also taken as a signature of critical slowing down \cite{Niermann2015}.

For symmetry-breaking phase transitions in a highly nonequilibrium setting, the dynamics are much less understood. Recent studies have found important features in nonequilibrium transitions, such as topological defects, which are absent in their equilibrium counterparts \cite{Yusupov2010,Mertelj2013,Zong2018}. Despite the differences, a slowdown in dynamics is thought to carry over to systems far from equilibrium. For example, in a rapid quench into a broken-symmetry state, the Kibble-Zurek theory suggests that critical slowing down plays a central role in domain formation: as the phase boundary is traversed at a faster rate than the system can respond, spatially disconnected regions may adopt distinct configurations of the same degenerate ground state \cite{Zurek1996}. Characteristic domain structures in liquid crystals have indeed been observed \cite{Chuang1991,Bowick1994}, providing indirect evidence for the slowdown. 

To study the dynamics in a nonequilibrium setting, charge-density-wave (CDW) transitions instigated by an intense femtosecond laser pulse provide an accessible platform with well-controlled tuning parameters. A suite of time-resolved probes can track the evolution of electronic and lattice orders after strong photoexcitation \cite{Zong2018}, offering insights into the critical behavior, if present, during the phase transition. Immediately after photoexcitation, a coherently-excited CDW amplitude mode was observed to soften transiently \cite{Yusupov2010}, hinting at critical slowing down. Right below $T_c$, a diverging relaxation time  back to equilibrium was interpreted as another signature \cite{Demsar2009,Zhu2018}. However, observables in previous studies, such as amplitude mode frequency or quasiparticle relaxation time across the spectroscopic gap, are only well defined in the broken-symmetry state \cite{Demsar1999, Kabanov1999,Yusupov2008}. To demonstrate slowing dynamics in the vicinity of a nonequilibrium phase transition, ideally one would measure an increased timescale near the phase boundary compared to both ordered and disordered states.

In this work, we circumvent this obstacle by focusing on a different observable during the photo-induced melting of a CDW: the time taken to suppress the condensate. With increasing photoexcitation densities, the perturbed system will enter one of the two transient states, where the CDW is either partially or completely suppressed \cite{Zong2018}. The two states are separated by the threshold excitation density, $F_\text{melt}$, where the condensate first vanishes. Through transient reflectivity and time-resolved diffraction measurements at different excitation densities, we observed that it takes the maximum time to suppress the CDW right at $F_\text{melt}$, indicating dynamical slowing down near the boundary between the two transient states. Here, we use \emph{dynamical} slowing down to emphasize the highly nonequilibrium nature of the system and to distinguish it from \emph{critical} slowing down commonly defined in a second-order phase transition in equilibrium \cite{Collins1989,Goldenfeld1992}.

\begin{figure}
    \includegraphics[scale=0.39]{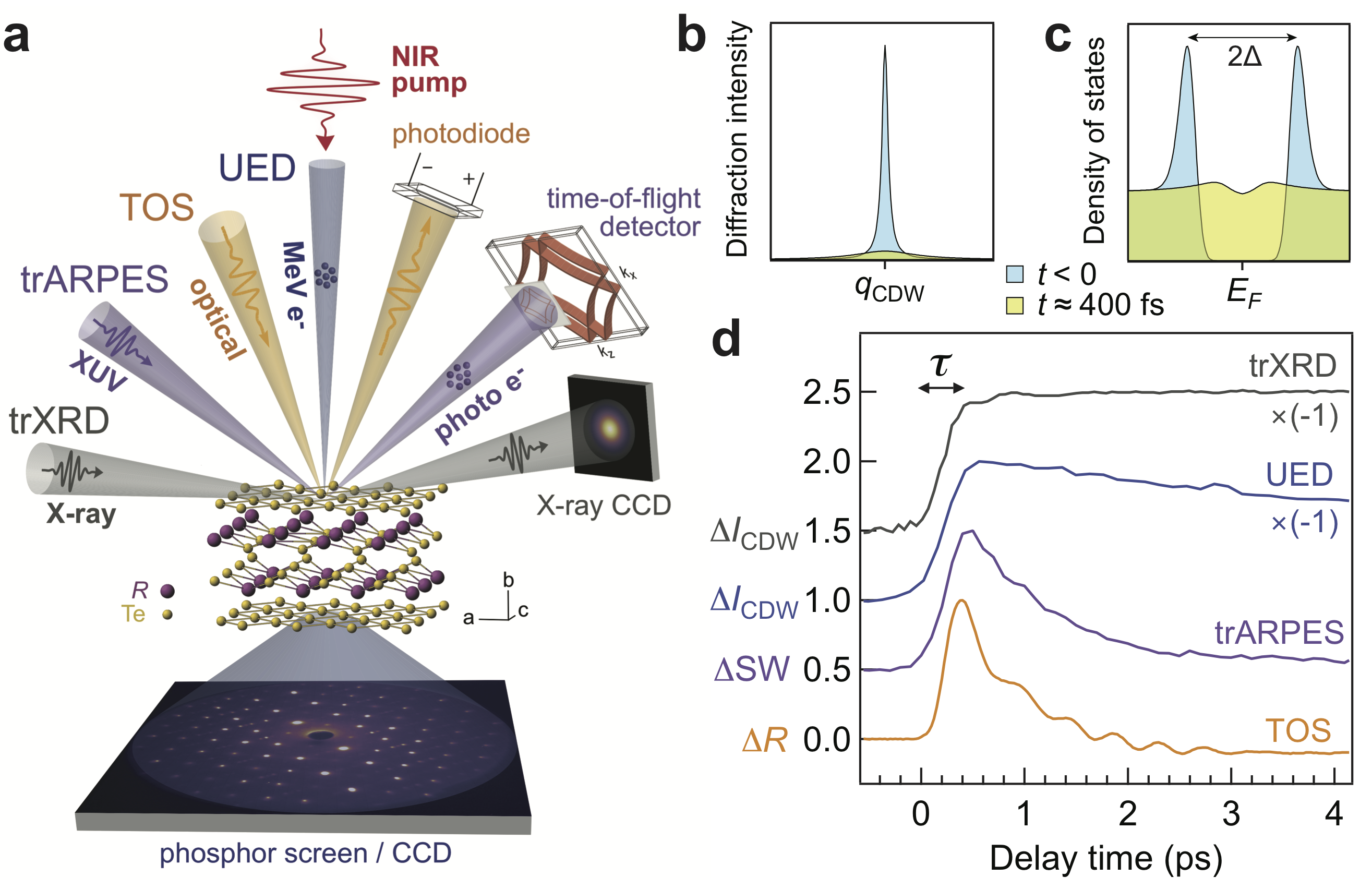}
    \caption{(Color online) Photo-induced CDW melting probed by multiple time-resolved techniques. (a)~Schematics of time-resolved probes, including ultrafast electron diffraction (UED), transient optical spectroscopy (TOS), time- and angle-resolved photoemission spectroscopy (trARPES), and time-resolved X-ray diffraction (trXRD). The full electron diffraction pattern is shown in Fig.\,\ref{fig:thz}(b). (b,c)~Schematics of the superlattice peak and density of states before (blue) and after (yellow) photoexcitation. (d)~Transient response of superlattice peak intensity ($\Delta I_\text{CDW}$), in-gap spectral weight ($\Delta$SW), and reflectivity ($\Delta R$) probed by corresponding time-resolved techniques. All traces are normalized between 0 and 1 and vertically offset for clarity. $\Delta I_\text{CDW}$ is inverted for easier comparison. All traces are measured in LaTe$_3$ except for trXRD, which measures TbTe$_3$, a similar compound in the same rare-earth tritelluride family with a lower $T_c$. The trace of trARPES is adapted from ref.~\cite{Zong2018}. The trace of trXRD is adapted with permission from ref.~\cite{Moore2016}; copyrighted by the American Physical Society.}
    \label{fig:intro}
\end{figure}

The material of interest is a paradigmatic CDW system, LaTe$_3$ \cite{Ru2008}. Like other rare-earth tritellurides, LaTe$_3$ possesses a quasi-2D structure (Fig.\,\ref{fig:intro}(a)) and develops a unidirectional CDW with wavevector $\mathbf{q}_\text{CDW}$ along the $c$-axis below $T_c\approx670$\,K \cite{Wang2014}. In equilibrium, the CDW transition is characterized by the appearance of satellite peaks in a diffraction pattern (Fig.\,\ref{fig:thz}(b)) as well as gap openings at certain parts of the Fermi surface connected by $\mathbf{q}_\text{CDW}$ \cite{Brouet2008}. 

Upon the arrival of a strong femtosecond laser pulse, the CDW order is transiently suppressed \cite{Schmitt2008,Moore2016,Zong2018}. We first establish the timescale for this process by performing ultrafast electron diffraction (UED) and transient optical spectroscopy (TOS), which reveal how the lattice and electrons respond to intense photoexcitation, respectively (Fig.\,\ref{fig:intro}(a)). Previous measurements from time- and angle-resolved photoemission spectroscopy (trARPES) \cite{Zong2018} and time-resolved X-ray diffraction (trXRD) \cite{Moore2016} are also included to obtain a comprehensive and consistent view of the ultrafast melting process.

\begin{figure}
    \includegraphics[scale=0.525]{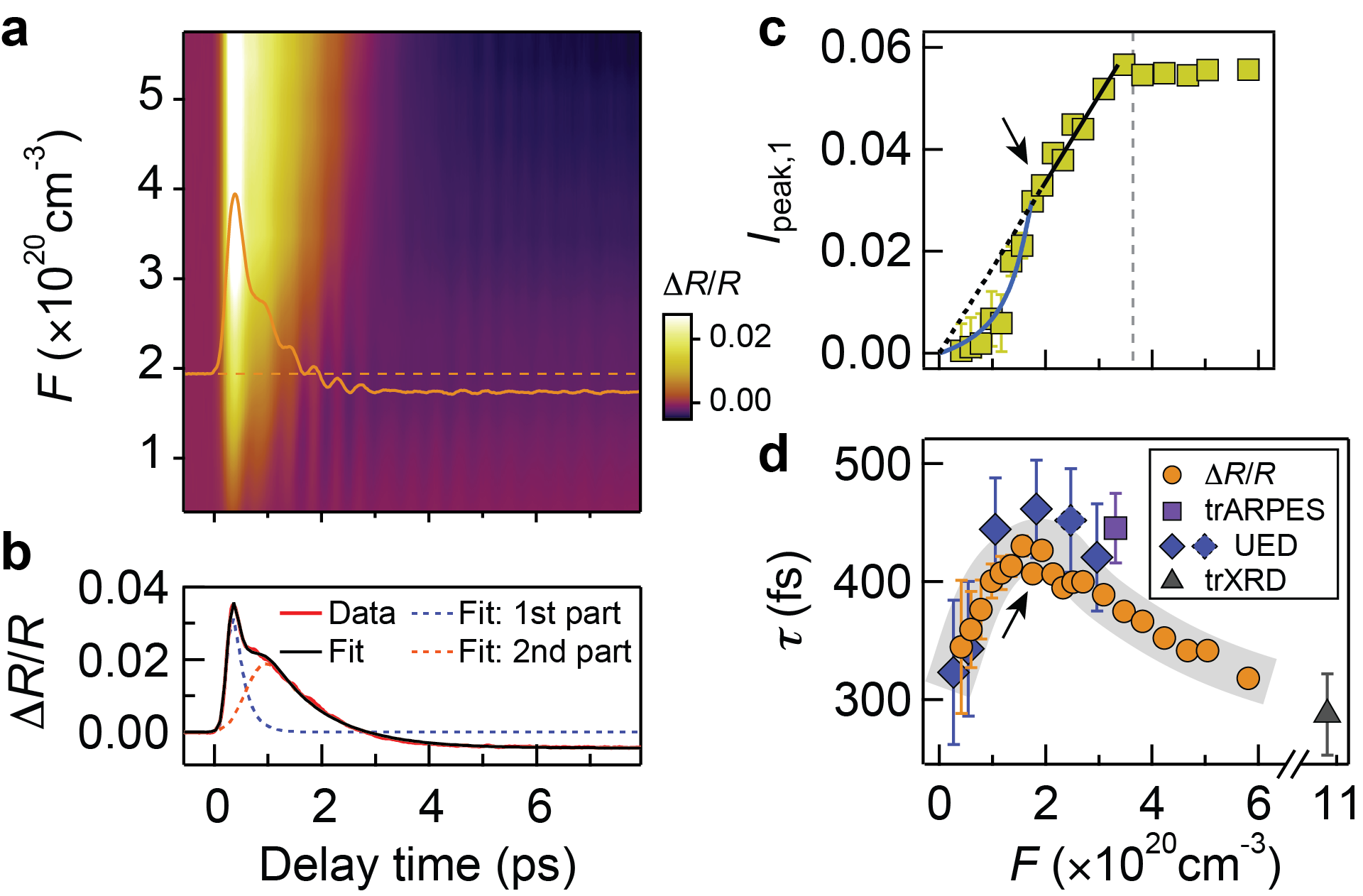}
    \caption{(Color online) CDW suppression time at different excitation densities. (a)~$\Delta R/R$ traces at different excitation densities, $F$, expressed in terms of absorbed photons per unit volume. A particular $\Delta R/R$ cut is overlaid at the excitation density indicated by the dashed line. It is the same trace shown in Fig.\,\ref{fig:intro}(d). (b)~$\Delta R/R$ trace at $F=5.81\times10^{20}$\,cm$^{-3}$, together with an example fit from Eq.\,\eqref{eq:pp_fit}. (c)~Initial excited quasiparticle population, $I_{\text{peak},1}$, showing a super-linear dependence on $F$ below $F_\text{melt}$ (arrow) and a plateau beyond $F_\text{bleach}$ (vertical dashed line). Black line is a linear fit with extrapolation (dashed) to zero. Blue curve is a fit to Eq.\,\eqref{eq:I_peak1}. (d)~CDW suppression time, $\tau$, as a function of excitation density, $F$. Gray curve is a guide to eye. For the UED data, the corresponding trace of the dashed diamond is shown in Fig.\,\ref{fig:intro}(d) while the rest, measured on a separate sample, are shown in Fig.\,\ref{fig:ued_flu_depend}. Error bars indicate uncertainties in curve fittings and in the instrumental temporal resolution \cite{SM}.}
    \label{fig:pp}
\end{figure}

While UED and trXRD track the evolution of CDW satellite peaks at characteristic wavevector $\mathbf{q}_\text{CDW}$ (Fig.\,\ref{fig:intro}(b)), TOS and trARPES probe the change in the spectroscopic gap (Fig.\,\ref{fig:intro}(c)) \cite{Zong2018}. Despite the different observables, the initial response that corresponds to CDW melting proceeds with a similar timescale, denoted by $\tau$ (Fig.\,\ref{fig:intro}(d)). The rising edges across the four techniques in Fig.\,\ref{fig:intro}(d) all span a time interval of $\tau\approx 400$\,fs, with variations arising from the different temporal resolutions in each setup \cite{SM} and different photoexcitation densities used (Fig.\,\ref{fig:pp}(d)). The agreement among structural and electronic probes suggests the presence of strong electron-phonon coupling in this system. Notably, the value of 400\,fs is on the same scale as the period of the 2.2\,THz CDW amplitude mode \cite{Yusupov2008}, further indicating the vital role of lattice vibrations in the formation of the charge order \cite{Hellmann2012}.

Among the four techniques discussed, TOS possesses the best temporal resolution and signal-to-noise ratio \cite{SM}, enabling us to more quantitatively investigate the timescale of CDW suppression, $\tau$, as we vary the laser excitation density, $F$, quoted in terms of absorbed photon number per unit volume \cite{Zong2018}. Figure~\ref{fig:pp}(a) shows the temporal evolution of the transient reflectivity, $\Delta R/R$, across a large range of $F$. The trace from Fig.\,\ref{fig:intro}(d) is overlaid at the corresponding $F$. The data presented was taken using a probe photon energy of 1.80\,eV (690\,nm), which is selected among the white light super-continuum because it is the energy most sensitive to the dynamics of the CDW gap \cite{SM}. In Fig.\,\ref{fig:pp}(b), we present an example cut (red curve) at $F=5.81\times10^{20}$\,cm$^{-3}$. To quantitatively evaluate the initial response time $\tau$, we performed a global fit for traces at all excitation densities using a two-component phenomenological model with minimal parameters. An example fit is presented in Fig.\,\ref{fig:pp}(b), showing excellent agreement. The two components arise from quasiparticle excitations in different parts of the Brillouin zone (see ref.~\cite{SM} for details of the fitting model and its interpretation), and we extracted $\tau$ from the rise time of the first component (Fig.\,\ref{fig:pp}(b), blue dashed curve).

Remarkably, the rise time $\tau$ displays a non-monotonic trend as a function of excitation density (Fig.\,\ref{fig:pp}(d), orange circles), with a maximum at $\sim2\times10^{20}$\,cm$^{-3}$ (black arrow). The non-monotonic trend of $\tau$ is independent of the fitting model, and is clearly observed in the raw data (Fig.\,\ref{fig:pp_more}(a)). To confirm this observation, we similarly track the suppression of superlattice peaks using UED at various excitation densities (Fig.\,\ref{fig:pp}(d), blue diamonds and Fig.\,\ref{fig:ued_flu_depend}). Despite significantly larger errors due to lower signal-to-noise ratio and poorer temporal resolution compared to the TOS measurements, the initial timescale in the UED experiments suggests the same non-monotonic behavior in $\tau$. We further note a recent measurement on SmTe$_3$ \cite{Trigo2018}, a CDW compound in the same family as LaTe$_3$, which demonstrates a similar trend in the initial system response.

To associate this non-monotonic behavior in $\tau$ with dynamical slowing down during the photo-induced CDW melting, we next establish that the melting proceeds the slowest precisely at the threshold excitation density when the CDW in the illuminated sample volume is just fully destroyed, namely, $F_\text{melt}\approx2\times10^{20}$\,cm$^{-3}$. We make three observations in this regard. First, the value $2\times10^{20}$\,cm$^{-3}$ corresponds to the point where the superlattice peak is observed to completely disappear in UED measurements \cite{Zong2018}, suggesting that $\tau$ indeed peaks at the threshold excitation density. Second, the time for the initial fast relaxation in transient reflectivity displays a steeply-increasing trend at $F_\text{melt}$ (Fig.\,\ref{fig:pp_more}(c)). This is attributed to a vanishing energy gap at the Fermi level when the CDW is completely suppressed, which limits the decay rate of excited quasiparticles \cite{Yusupov2008,Demsar1999,Kabanov1999}. Third, the maximum reflectivity change, $I_{\text{peak},1}$, also displays distinct behavior below and above $F_\text{melt}$ (Fig.\,\ref{fig:pp}(c)). Below $F_\text{melt}$, the presence of a CDW gap modifies the transient population of excited quasiparticles, resulting in a super-linear $I_{\text{peak},1}$ as a function of excitation density (Fig.\,\ref{fig:pp}(c), blue curve; see \cite{SM}). Beyond $F_\text{melt}$, the excited quasiparticle population is directly proportional to the excitation density (Fig.\,\ref{fig:pp}(c), black line). Above an even higher value $F_\text{bleach}$, the peak reflectivity $I_{\text{peak},1}$ plateaus (Fig.\,\ref{fig:pp}(c), vertical dashed line) due to quasiparticle bleaching \cite{SM}. It is worth emphasizing that at $F_\text{melt}$, the lattice temperature stays below $T_c$ at all time delays after photoexcitation \cite{Zong2018}, reaffirming that the observed CDW melting is non-thermal in nature.
 
To interpret the non-monotonic trend of the initial response time ($\tau$) measured in TOS, we need to understand what physical quantity is probed by transient reflectivity. Unlike the superlattice peak intensity or in-gap spectral weight, optical reflectivity is not a direct gauge of the CDW order parameter. Typically, in a gapped system, the value of transient reflectivity is taken to be proportional to the excited quasiparticle density \cite{Yusupov2008,Demsar1999,Kabanov1999}, which in turn is sensitive to the gap size. For example, clear oscillations are present in $\Delta R$ traces (Figs.\,\ref{fig:intro}(d), \ref{fig:pp}(a), and \ref{fig:pp_wl}), with a dominant contribution from the CDW amplitude mode \cite{Yusupov2008} -- the modulation of the gap magnitude. Based on this sensitivity of $\Delta R$ to the gap size as well as the consistency of the initial timescale in Figs.\,\ref{fig:intro}(d) and \ref{fig:pp}(d) across techniques, we take the initial rise time ($\tau$) in transient reflectivity as the time needed for the amplitude of the CDW order parameter to be maximally suppressed. The value of $\tau$ is well separated from any electron-electron scattering timescale ($\leq100$\,fs) \cite{Demsar1999,Prasankumar2016}, and represents a simultaneous population of excited quasiparticles and renormalization of the gap, which occur self-consistently.

Having established the precise meaning of $\tau$, we draw some parallels between the present nonequilibrium study and its equilibrium counterparts to interpret the observation in Fig.\,\ref{fig:pp}(d). At equilibrium, when the temperature is close to $T_c$, time-domain measurements of the order parameter indicate a reduced rate of change, which signifies critical slowing down \cite{Collins1973,Iizumi1986,Horie1987}. Here, we use photoexcitation density in lieu of temperature as the tuning parameter, and we extend the timescale to the femtosecond regime. Similarly, we interpret the maximum value of $\tau$ at exactly the threshold excitation density as a signature of dynamical slowing down in this ultrafast phase transition.

To understand how a slowdown in dynamics can be extended to a regime far from equilibrium, we again make reference to the established framework of symmetry-breaking transition in equilibrium, which is parameterized by an order parameter $\psi$. On a phenomenological level, we consider the standard Landau potential \cite{SM}, ${\cal W}(\psi)$, which gives the simplest description of the second-order CDW transition in LaTe$_3$ \cite{Ru2008}. To see the slowdown near $T_c$, the usual treatment is to solve the time-dependent Landau equation \cite{Goldenfeld1992}, $\partial \psi/\partial t = -\Gamma \delta {\cal W}/\delta \psi$, where $\Gamma$ is a phenomenological parameter. Close to $T_c$ where the order parameter $\psi$ is small and the free energy ${\cal W}(\psi)$ develops a flat bottom, the relaxation time of $\psi$ after any small perturbation can be shown to approach infinity, which is the origin of critical slowing down \cite{Goldenfeld1992}.

\begin{figure}
    \includegraphics[scale=0.6]{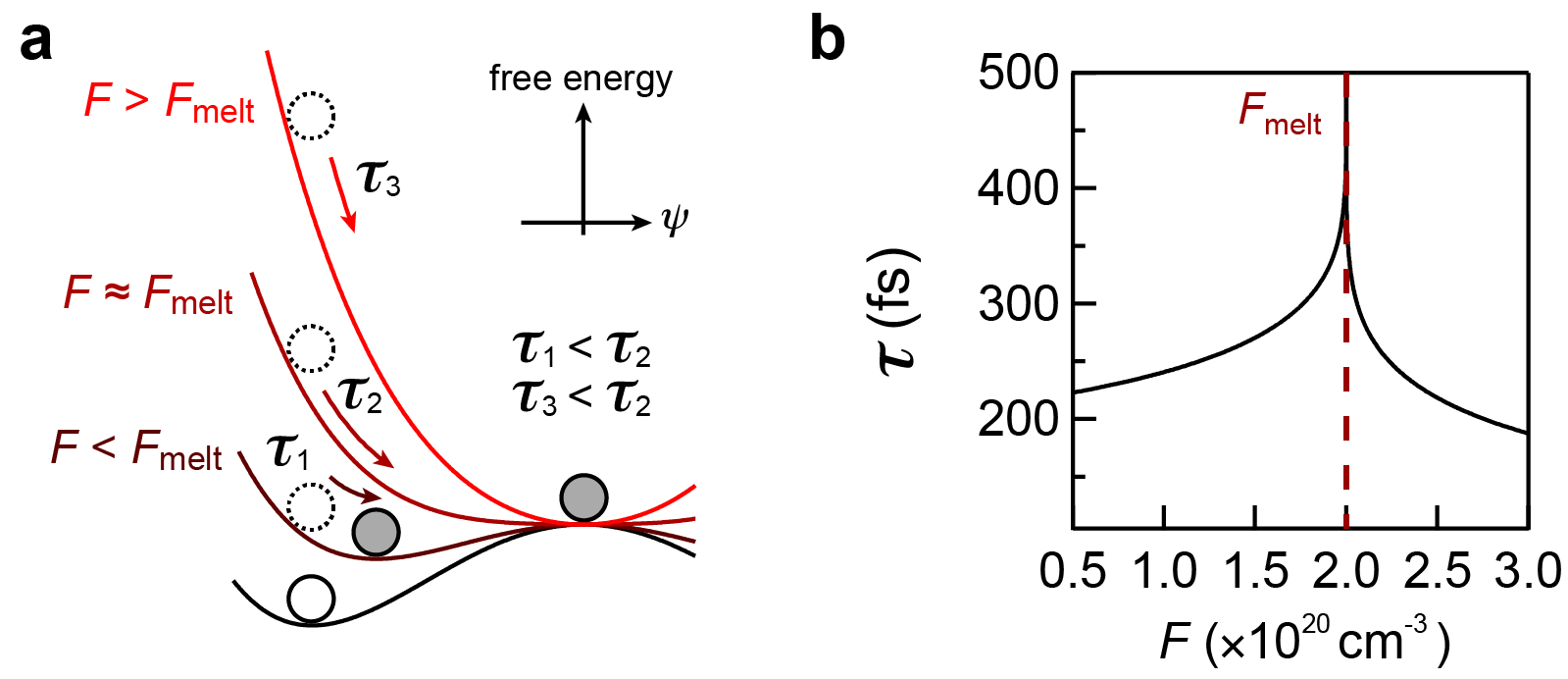}
    \caption{(Color online) Dynamical slowing down in the generalized time-dependent Landau theory. (a)~Schematic of CDW order parameter ($\psi$) dynamics in a Landau free energy landscape. The solid circle represents $\psi$ before photoexcitation and dashed ones are nonequilibrium $\psi$ in an impulsively altered free energy, whose subsequent evolution, which is not drawn, is described by Eq.\,\eqref{eq:alpha}. Filled circles represent $\psi$ when CDW is transiently suppressed, either partially or completely. Colored curves are snapshots of transient free energies at different excitation densities after laser pulse incidence. At $F_\text{melt}$, $\psi$ first reaches zero in its temporal evolution. (b)~Calculated CDW suppression time as a function of excitation density, using the time-dependent Landau theory \cite{SM}.}
    \label{fig:theory}
\end{figure}

We generalize this treatment to a highly nonequilibrium situation by considering an impulsive change to the free energy ${\cal W}(\psi)$ that mimics the photoexcitation. To account for the different responses by the electronic and phononic subsystems, we further consider two components, one for the electrons and the other for the lattice. In our model, the two components are strongly coupled, as supported by the observation of a similar melting timescale $\tau$ across different probes (Fig.\,\ref{fig:intro}(d)). Details of the calculation are described in ref.\,\cite{SM}; here, we only highlight the physical picture summarized in Fig.\,\ref{fig:theory}(a). The laser pulse significantly modifies the free energy ${\cal W}(\psi)$, setting off the order parameter to seek a new global minimum. Though we draw ${\cal W}(\psi)$ as fixed curves, it should be noted that the free energy evolves dynamically according to Eq.\,\eqref{eq:alpha}. At the critical excitation density, $F_\text{melt}$, beyond which the order parameter vanishes transiently, the time taken to suppress the order is the longest ($\tau_2 > \tau_1, \tau_3$ in Fig.\,\ref{fig:theory}(a)). Similar to the equilibrium situation, the slow evolution reflects a transiently flat potential landscape when the order parameter is close to zero, which leads to its reduced rate of change.

Using our experimental parameters for the time-dependent Landau equation, the calculated CDW suppression time, $\tau$, is shown in Fig.\,\ref{fig:theory}(b) (see ref.~\cite{SM} for details). There is no adjustable parameter except a constant that converts a dimensionful $F$ in the experiment to a dimensionless quantity in the computation. Here, $\tau$ is defined as the time spanned between the arrival of the laser pulse and the transient minimum position of $|\psi|^2$. It shows a distinct peak at the critical point, $F_\text{melt}$, which captures the experimental observation in Fig.\,\ref{fig:pp}(d). Notably, the absolute value of the calculated $\tau$ falls under a similar range of magnitudes as observed in experiment. This timescale is determined by the period of the CDW amplitude mode in the simulation \cite{SM}, indicating the instrumental role of phonons in mediating the ultrafast transition.

There is one key difference between the calculated and measured trend of $\tau$: the latter lacks a sharp divergence at the threshold excitation density. We attribute this rounding of the divergence to the presence of topological defects \cite{Goldenfeld1992}, which are known to exist after photoexcitation \cite{Yusupov2010,Mertelj2013,Zong2018}. In the Landau picture, they disrupt the local gradient in the free energy, avoiding the divergence that requires a flat energy landscape in spatially extended regions. Furthermore, the divergence only happens in a very narrow window of excitation densities (Fig.\,\ref{fig:theory}(b)), which makes experimental detection challenging as any small uncertainties or fluctuations in the pulse energy can smear the singularity.

In conclusion, two different time-resolved probes are used to systematically study the ultrafast melting of a CDW instigated by an intense laser pulse. We have experimentally demonstrated the phenomenon of dynamical slowing down, manifested as the longest time it takes to suppress the CDW at the threshold excitation density in the nonequilibrium phase transition. The agreement in timescale across techniques and with theoretical simulation by time-dependent Landau equations highlights the important role of phonons in this photo-induced transition. Despite complexities involved in phase transitions far from equilibrium, the observation of slowing dynamics in this setting pinpoints a robust commonality for us to understand nonequilibirum phenomena of more intricate systems.

\begin{acknowledgments}
We acknowledge helpful discussions with B.V.~Fine, A.V.~Rozhkov, and E.~Baldini. We thank E.J.~Sie for assistance in sample handling. We acknowledge support from the U.S. Department of Energy, BES DMSE (experimental setup and data acquisition), from the Gordon and Betty Moore Foundation's EPiQS Initiative grant GBMF4540 (data analysis and manuscript writing), and the Skoltech NGP Program (Skoltech-MIT joint project) (theory). I.T. and H.W. acknowledge the support from the U.S. Department of Energy, Office of Science, Office of Basic Energy Sciences, under Contract No. DE-SC0012509 (data acquisition). Y.-Q.B. was supported by the Center for Excitonics, an Energy Frontier Research Center funded by the U.S. Department of Energy, Office of Science, Office of Basic Energy Sciences, under award No. DESC0001088 (sample characterization). Y.Y. was supported by the Center for Integrated Quantum Materials under NSF grant DMR-1231319 (device fabrication). Research in the P.J.-H. group (Y.-Q.B., Y.Y., X.W., and P.J.-H.) was partly supported by the Gordon and Betty Moore Foundation's EPiQS Initiative through grant GBMF4541 (sample preparation and characterization). J.S. and I.R.F. acknowledge support from the U.S. Department of Energy, Office of Basic Energy Sciences, under Contract No. DE-AC02-76SF00515 (sample growth and characterization). X.S., R.L., J.Y., S.P., M.C.H., B.K.O.-O., M.E.K., and X.W. acknowledge support by the U.S. Department of Energy BES SUF Division Accelerator \& Detector R\&D program, the LCLS Facility, and SLAC under contract No.'s DE-AC02-05-CH11231 and DE-AC02-76SF00515 (MeV UED). 
\end{acknowledgments}

%

\clearpage
\newpage
\beginsupplement
\section{Supplemental Materials}

\subsection{I. Details of time-resolved techniques}

\begin{figure}
    \includegraphics[scale=0.395]{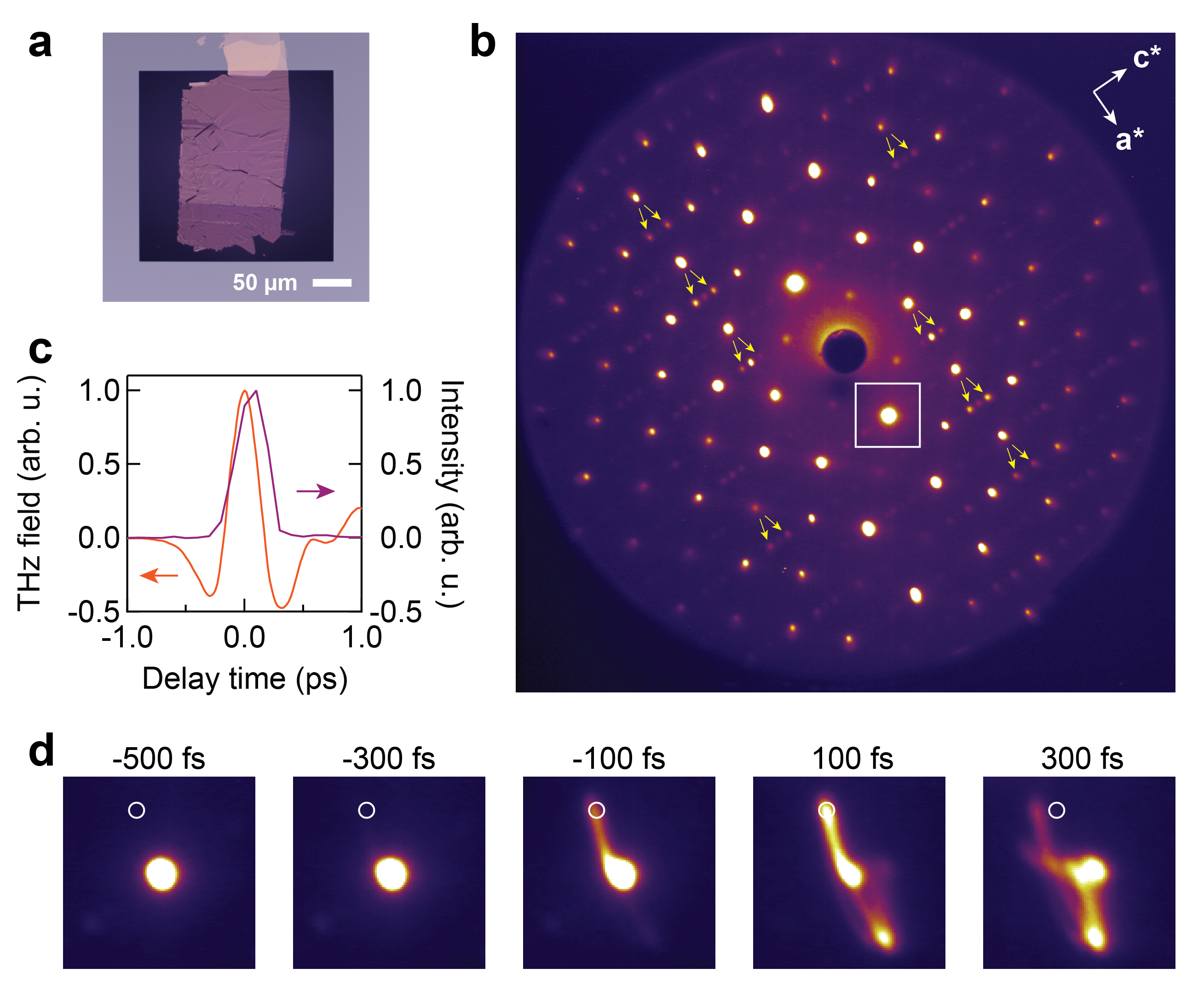}
    \caption{(Color online) UED and THz streaking. (a)~Optical image of the UED sample, exfoliated in an inert gas environment onto a 10\,nm-thick silicon nitride window. (b)~Full room-temperature diffraction pattern of LaTe$_3$ in the (H~0~L) plane with 3.1\,MeV incident electron kinetic energy. Pairs of arrows indicate selected superlattice peaks whose intensity averages yield the curve shown in Fig.\,\ref{fig:intro}(d). These peaks are selected for best signal-to-noise as their intensity exceeds a preset threshold. White square denotes the region of interest (ROI) of the (2~0~0) Bragg peak in (d). (c)~THz field strength $E(t)$ (left) from electro-optic sampling and intensity $I(t)$ (right) integrated over the circular ROI shown in (d). Both quantities are normalized for easier comparison. The initial rise of $I(t)$ results from THz-induced electron deflection. (d)~Snapshots of the (2~0~0) Bragg peak at selected time delays, showing electron deflection due to the THz field.}
    \label{fig:thz}
\end{figure}

\textbf{MeV ultrafast electron diffraction (UED).} The experiment was carried out on an exfoliated thin flake of LaTe$_3$ (Fig.\,\ref{fig:thz}(a)) at room temperature, prepared using the same method described in ref.~\cite{Zong2018}. The measurement was performed in the MeV UED setup in the Accelerator Structure Test Area facility at SLAC National Laboratory \cite{Weathersby2015,Shen2018} using a 3.1\,MeV electron bunch operating at 180\,Hz. The electron beam size on the sample is approximately 90\,$\upmu$m by 90\,$\upmu$m (full-width at half maximum, or FWHM), much smaller than the 800\,nm (1.55\,eV) pump laser beam from a commercial Ti:sapphire laser (Vitara and Legend Elite HE, Coherent Inc.). 

At this pump laser wavelength, the penetration depth ($1/e$ of intensity) is 44\,nm, calculated from static reflectivity and optical conductivity in ref.~\cite{Sacchetti2006, Pfuner2009}. This depth is comparable to the sample thickness, so not all layers are uniformly photoexcited even though all layers equally contribute to the electron diffraction pattern at 3.1\,MeV beam energy. We therefore estimate an effective excitation density $F$ based on the recovery timescale of the superlattice peak, which is shown to have an approximately linear dependence on $F$ in the range probed \cite{Zong2018}.

\textbf{Transient optical spectroscopy (TOS).} The experiment was carried out at MIT using the output of a commercial Ti:sapphire laser (Wyvern 500, KMLabs) operating at 30\,kHz. The laser beam was split into a pump branch (780\,nm, 1.59\,eV) and a probe branch, with the latter focused to a sapphire crystal to generate a white light continuum (500\,nm to 700\,nm; 2.48\,eV to 1.77\,eV). Both branches were focused onto a freshly cleaved sample in the (010) plane held at room temperature at near-normal incidence and parallel polarization. The reflected probe beam was directed to a monochromator and photodiode for lock-in detection. As determined from the pump-probe cross-correlation, the overall temporal resolution is 70\,fs.

The data presented in the main text selects a probe wavelength of 690\,nm (1.80\,eV) as it is the wavelength most sensitive to the CDW gap dynamics in the spectral range probed (see Sec.\,V). 

\subsection{II. Determining UED temporal resolution by terahertz streaking}

The investigation of the initial system response after photoexcitation requires the knowledge of the instrumental temporal resolution. Unlike transient optical spectroscopy where the resolution can be extracted from a pump-probe cross-correlation experiment, it is less straightforward to measure in UED. One common method is to perform UED on a reference sample, where the initial dynamics is known to be fast \cite{Weathersby2015}. However, if the temporal resolution is comparable to the sample response time, a sample-independent method is preferred to disentangle the resolution effect from the intrinsic response.

For this purpose, we determine the temporal resolution by streaking the electron bunch in an intense terahertz (THz) field, from which the electron profile can be extracted \cite{RKLi2018,Ofori-Okai2018}. The THz pulse
($\sim$\,650\,kV/cm) was generated by a DSTMS (4-N,N-dimethylamino-4'-N'-methyl-stilbazolium 2,4,6-trimethylbenzenesulfonate) crystal with the spectral peak at 2\,THz. A typical THz electric field profile, $E(t)$, is shown in Fig.\,\ref{fig:thz}(c) (left), measured via electro-optic sampling with an 800\,nm (1.55\,eV), 80\,fs pulse. As the single-cycle THz pulse and the electron bunch overlap spatially and temporally on the sample, the electrons are deflected sideways due to the strong net Lorentz force (Fig.\,\ref{fig:thz}(d)) \cite{RKLi2018}. Therefore, one can estimate the instrumental temporal resolution, in this case dominated by the contribution from the electron source, by observing the streaking pattern.

More specifically, as the THz pulse is incident on the sample, we monitor the temporal evolution of integrated intensity, $I(t)$, of a circular region of interest (ROI), whose diameter is the FWHM of the nearby Bragg peak (Fig.\,\ref{fig:thz}(c) right and (d)). The finite width of the rising edge in $I(t)$ gives an upper bound of the electron bunch length, as the rising edge has additional contributions from a nonzero transverse bunch size and a finite streaking speed. It should be noted that we only use the initial rising edge of $I(t)$. This is because motions of the MeV electrons after 100\,fs are more complex when electrons inside the metallic sample interact with the THz pulse and distort the local electric field (Fig.\,\ref{fig:thz}(d)). As the streaked electrons can span the entire detector segment from its original position to the extremum position in the ROI (Fig.\,\ref{fig:thz}(d)), the bunch width is therefore lower-bounded by the width of the rising edge in the THz pulse, $E(t)$ (Fig.\,\ref{fig:thz}(c), left). By fitting the rising edges of $E(t)$ and $I(t)$ to an erf-function, we determine the UED temporal resolution from the fitted FWHM to be between 280\,fs and 325\,fs. This uncertainty in the temporal resolution is translated to the corresponding error bars in Fig.\,\ref{fig:pp}(d).

\subsection{III. Fitting and interpretation of time traces}

\begin{figure}[ht!]
    \includegraphics[scale=0.48]{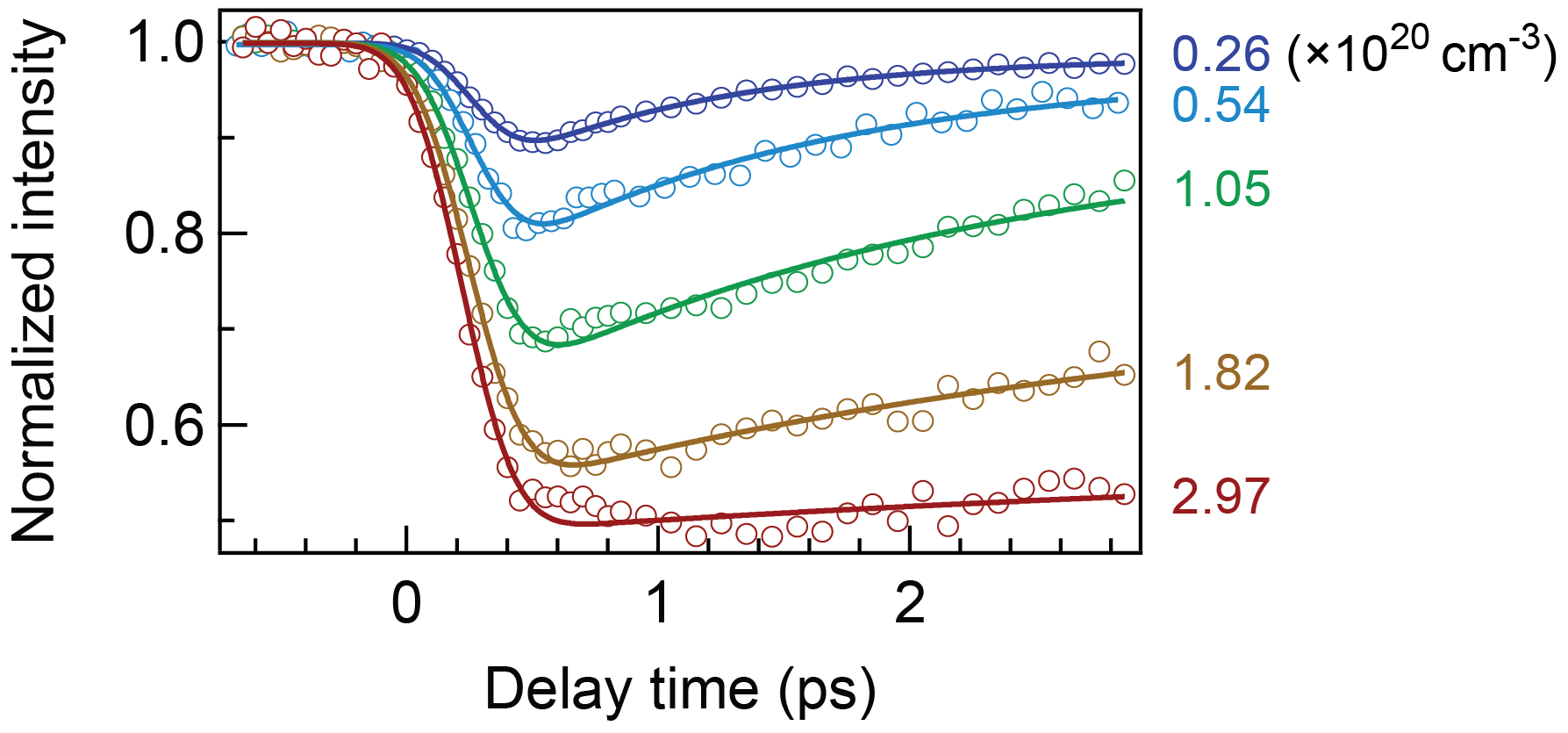}
    \caption{(Color online) Initial suppression of the superlattice peak intensity at different excitation densities, quoted in terms of absorbed photons per unit volume. Intensities are normalized to values before photoexcitation. Curves are fits to the raw data (see text). The non-vanishing intensity beyond the melting threshold arises from background intensity in the diffraction pattern and non-uniform illumination of all layers of the sample due to a finite pump laser penetration depth.}
    \label{fig:ued_flu_depend}
\end{figure}

For temporal evolutions measured in UED and for data in trARPES \cite{Zong2018} and trXRD \cite{Moore2016}, we use the typical erf-function with a single-exponential decay \cite{Moore2016,Zong2018} to fit the time traces shown in Fig.\,\ref{fig:intro}(d), where the CDW melting time is extracted from the FWHM of the erf-function. Example fits of the UED data are shown in Fig.\,\ref{fig:ued_flu_depend} at different excitation densities.

\begin{figure}
    \includegraphics[scale=0.52]{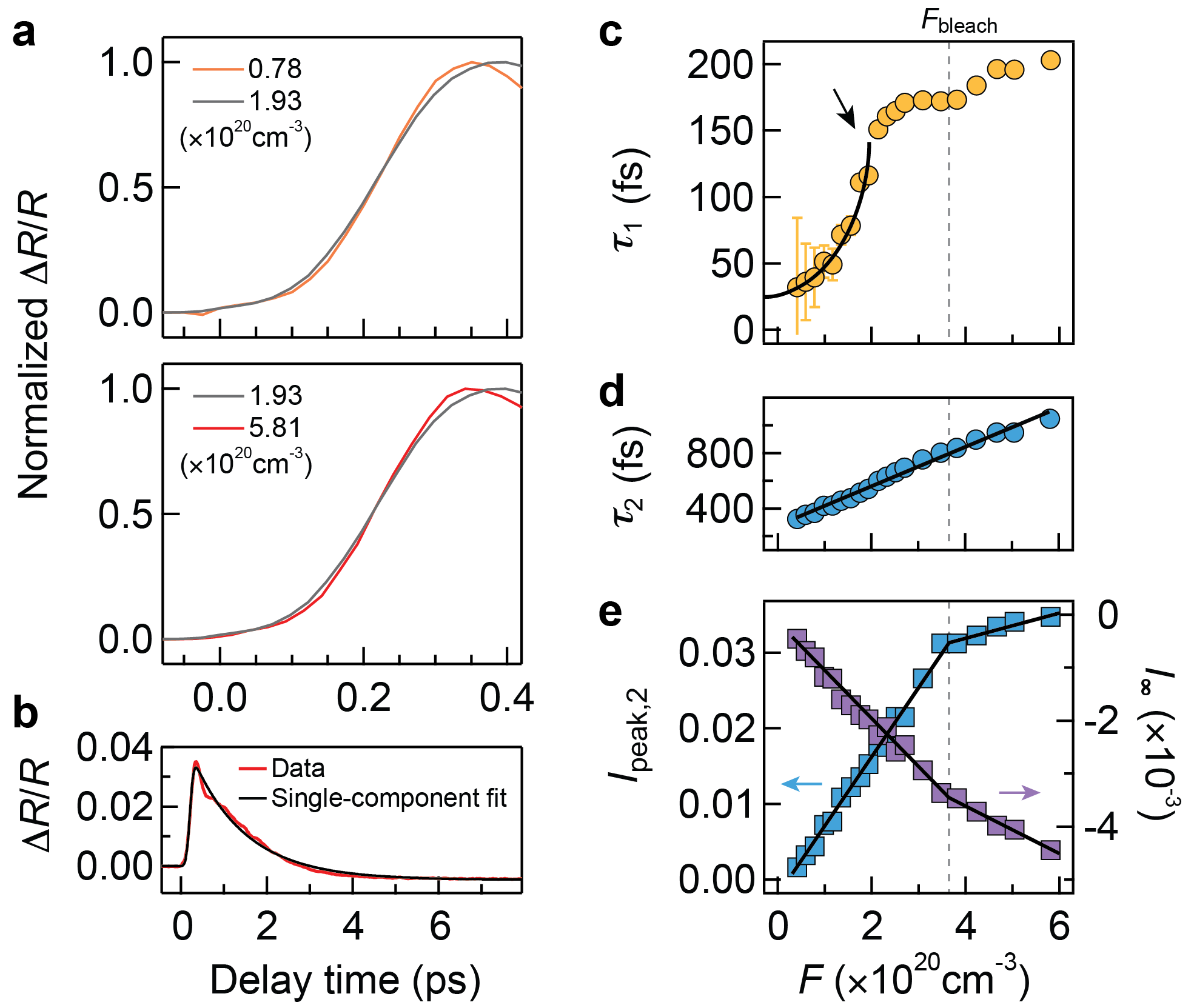}
    \caption{(Color online) CDW melting and quasiparticle dynamics measured by TOS. (a)~Selected traces of $\Delta R/R$ in the CDW melting time window, featuring a maximum melting timescale as $F$ reaches the threshold value of $F_\text{melt} \approx 2\times10^{20}$\,cm$^{-3}$, as exemplified by a less steep slope of the initial rise. All traces are normalized between 0 and 1 for easier comparison. (b)~$\Delta R/R$ trace at $F=5.81\times10^{20}$\,cm$^{-3}$, the same as shown in Fig.\,\ref{fig:pp}(b), but with a single-component fit instead, showing poor agreement from 1 to 4\,ps. (c)~Relaxation time of excited quasiparticles across the CDW gap ($\tau_1$), showing a steeply-increasing trend near $F_\text{melt}$ (black arrow), an indicator of CDW gap closing. Black curve is a guide to eye. (d)~Relaxation of excited quasiparticles in the ungapped part of the Brillouin zone ($\tau_2$), showing a linear increase with excitation density. (e)~Excited quasiparticle population in the ungapped momentum region, represented by $I_{\text{peak},2}$ (left), and lattice heating, represented by $I_\infty$ (right), both showing a kink at $F_\text{bleach}$ without any salient features at $F_\text{melt}$. Error bars, if larger than the marker size, represent uncertainties in the curve fitting. See Eq.\,\eqref{eq:pp_fit} for details.}
    \label{fig:pp_more}
\end{figure}

For transient reflectivity, we first applied the same single-exponential function. However, the fitting shows poor agreement at high excitation density, as exemplified in Fig.\,\ref{fig:pp_more}(b). In particular, the kink around 0.6\,ps in the $\Delta R/R$ trace leads to significant discrepancies between the data and the fitted curve from 1 to 4\,ps. To improve the agreement, we hence adopt the following phenomenological model with two components,
\begin{widetext}
\begin{align}\label{eq:pp_fit}
    \Delta R(t)/R(t) = \Bigg\{&\left[ \frac{1}{2}\left(1+\text{Erf}\left(2\sqrt{2}(t-t_0)/\tau\right)\right) I_{\text{peak},1}\,e^{-(t-t_0)/\tau_1}\right] \nonumber\\
    +& \left[ \frac{1}{2}\left(1+\text{Erf}\left(2\sqrt{2}(t-t_1)/\tau'\right)\right)\left(I_\infty+ I_{\text{peak},2}\,e^{-(t-t_1)/\tau_2}\right)\right] \Bigg\} * g(w_0,t).
\end{align}
\end{widetext}

Before discussing the meaning of the fitting parameters, we first describe the procedure and present the result. All traces shown in Fig.\,\ref{fig:pp}(a) are included in a global fitting algorithm, where $\tau'$ and $t_1$ are constrained to be the same for all excitation densities to limit the number of free parameters. Their fitted values are $\tau' = 1.078\pm0.007$\,ps and $t_1=0.763\pm0.003$\,ps. Other important parameters as a function of excitation density are shown in Figs.\,\ref{fig:pp}(c,d) and \ref{fig:pp_more}(c--e). For all fittings, we use a Gaussian kernel, $g(w_0,t)$, to model the intrinsic instrumental temporal resolution. The FWHM of the Gaussian, $w_0$, is 70\,fs for TOS, 230\,fs for trARPES \cite{Zong2018}, and 300\,fs for trXRD \cite{Moore2016}. The temporal resolution of the UED experiment has been discussed in Sec.~II.

To understand the physical picture behind the two components, which are represented by the two $[\,\cdot\,]$ terms in Eq.\,\eqref{eq:pp_fit}, we first compare their relaxation times, $\tau_1$ and $\tau_2$, shown in Fig.\,\ref{fig:pp_more}(c,d). While $\tau_1$ shows an abrupt change at $F_\text{melt}$, $\tau_2$ is featureless across the threshold excitation density. This suggests that the first component is more sensitive to the CDW gap, as the order parameter transiently vanishes at $F_\text{melt}$. Indeed, the steeply-increasing trend highlighted by the black curve in Fig.\,\ref{fig:pp_more}(c) is consistent with previous reports of suppressed quasiparticle relaxation in similar CDW systems close to $T_c$ \cite{Yusupov2008,Demsar1999,Kabanov1999}. Given that the Fermi surface is only partially gapped in the CDW state and the gap size is highly anisotropic in the Brillouin zone \cite{Brouet2008}, we associate the first $[\,\cdot\,]$ term with quasiparticle dynamics in the gapped region, while the second with that in the metallic region. We note that the smoothly-increasing $\tau_2$ as a function of excitation density is consistent with phonon-mediated quasiparticle relaxation in metallic systems \cite{Prasankumar2016}.

This interpretation of $\Delta R/R$ traces is further supported by the comparison between $I_\text{peak,1}$ (Fig.\,\ref{fig:pp}(c)) and $I_\text{peak,2}$ (Fig.\,\ref{fig:pp_more}(e), left). At the threshold $F_\text{melt}$, $I_\text{peak,1}$ changes from a super-linear to a linear trend, while $I_\text{peak,2}$ displays no distinctive feature. The super-linear trend in $I_\text{peak,1}$ is consistent with excited quasiparticle population in a gapped band structure, where the transient gap size is dependent on the excitation density (see Sec.~IV). On the other hand, the quasiparticle dynamics in the ungapped part of the Brillouin zone is less affected by the changing CDW gap, so there is a lack of feature in $I_\text{peak,2}$ at $F_\text{melt}$. 

With this interpretation in mind, our primary subject of the current study, $\tau$, represents the time to modify the excited quasiparticle population in the gapped region. Quantitatively, $\tau$ in the TOS measurement is consistent with the initial response time in other time-resolved techniques (Figs.\,\ref{fig:intro}(d) and \ref{fig:pp}(d)). Therefore, gap renormalization and quasiparticle excitation in the gapped region occur self-consistently. Compared to $\tau$, the rise time of the second component, $\tau'$, is considerably longer. This slower rise is accompanied by the relaxation of the first component (Fig.\,\ref{fig:pp}(b)), suggesting quasiparticle scattering across different parts of the Brillouin zone.

Another important parameter, $I_\infty$, represents the transient reflectivity at long delay time relative to the probed time window, indicating laser-induced heating. By plotting $I_\infty$ and $I_{\text{peak,}i=1,2}$ as a function of photoexcitation density (Figs.\,\ref{fig:pp}(c) and \ref{fig:pp_more}(e)), we further observe a kink at $F\approx3.6\times10^{20}$\,cm$^{-3}$. This value of excitation density corresponds to approximately 0.2 absorbed photons per unit cell of LaTe$_3$. We interpret this value as $F_\text{bleach}$, where bands near the Fermi level within the energy of the probe photon are depleted. It thus accounts for the plateau feature in $I_\text{peak,1}$ (Fig.\,\ref{fig:pp}(c)), where the population of excited quasiparticles is saturated.

\subsection{IV. Quasiparticle population across the gap below the critical excitation density}

In the main text, we identify the critical excitation density, $F_\text{melt}$, with the location of a kink in the maximum population of excited quasiparticles in the gapped region, $I_{\text{peak},1}$ (Fig.\,\ref{fig:pp}(c)). This is the threshold point where a super-linear trend of $I_{\text{peak},1}$ turns into a linear behavior, evident in the fitted blue curve and the black line in Fig.\,\ref{fig:pp}(c). In this section, we expand the discussion on the super-linear trend below $F_\text{melt}$.

At low excitation density where the CDW is not fully suppressed, a phonon-mediated decay channel of quasiparticles results in a maximum quasiparticle population of \cite{Yusupov2008,Kabanov1999}
\begin{equation}\label{eq:I_peak1}
    I_{\text{peak},1}=I_0 F\frac{(\Delta+k_B T/2)^{-1}}{1+\gamma\sqrt{2k_BT/\pi\Delta}\,e^{-\Delta/k_BT}},
\end{equation}
where $I_0$ is a constant of proportionality, $F$ is the excitation density, $k_B$ is the Boltzmann constant, $T=300$\,K is the sample temperature, and $\Delta$ is the CDW gap size. Here, $\gamma$ is a material specific parameter accounting for the phonon modes that participate in the relaxation of the quasiparticles \cite{Yusupov2008,Kabanov1999}. Strictly speaking, Eq.\,\eqref{eq:I_peak1} only applies to a system with an isotropic gap \cite{Kabanov1999}, but empirically it works well for the rare-earth tritelluride series where the gap size is momentum-dependent \cite{Yusupov2008}.

In the fit of Fig.\,\ref{fig:pp}(c) at $F<F_\text{melt}$, we assume the simplest case where the gap size $\Delta$ depends on $F$ linearly,
\begin{equation}\label{eq:delta}
    \Delta=\Delta_0(1-F/F_\text{melt}),
\end{equation}
where $\Delta_0/k_B = 4062$\,K is the equilibrium value of the gap at 300\,K \cite{Wang2014}. The only free parameters in the curve fitting of Fig.\,\ref{fig:pp}(c) are $I_0$ and $\gamma$, where the fitted value of $\gamma=2.8\pm0.5$. The fit reproduces the super-linear trend, in particular near $F=0$. When $F$ approaches $F_\text{melt}$, Eq.\,\eqref{eq:I_peak1} is no longer applicable as it predicts a vanishing $I_{\text{peak},1}$ as $\Delta$ vanishes. The discrepancy arises from the assumption in Eq.\,\eqref{eq:I_peak1} that $\Delta$ is constant in time while in the present study, $\Delta$ evolves as a function of time. Therefore, the fit presented in Fig.\,\ref{fig:pp}(c) is restricted to a range strictly below $F_\text{melt}$.

\subsection{V. Probe photon energy in transient optical spectroscopy}

\begin{figure}[ht!]
    \includegraphics[scale=0.54]{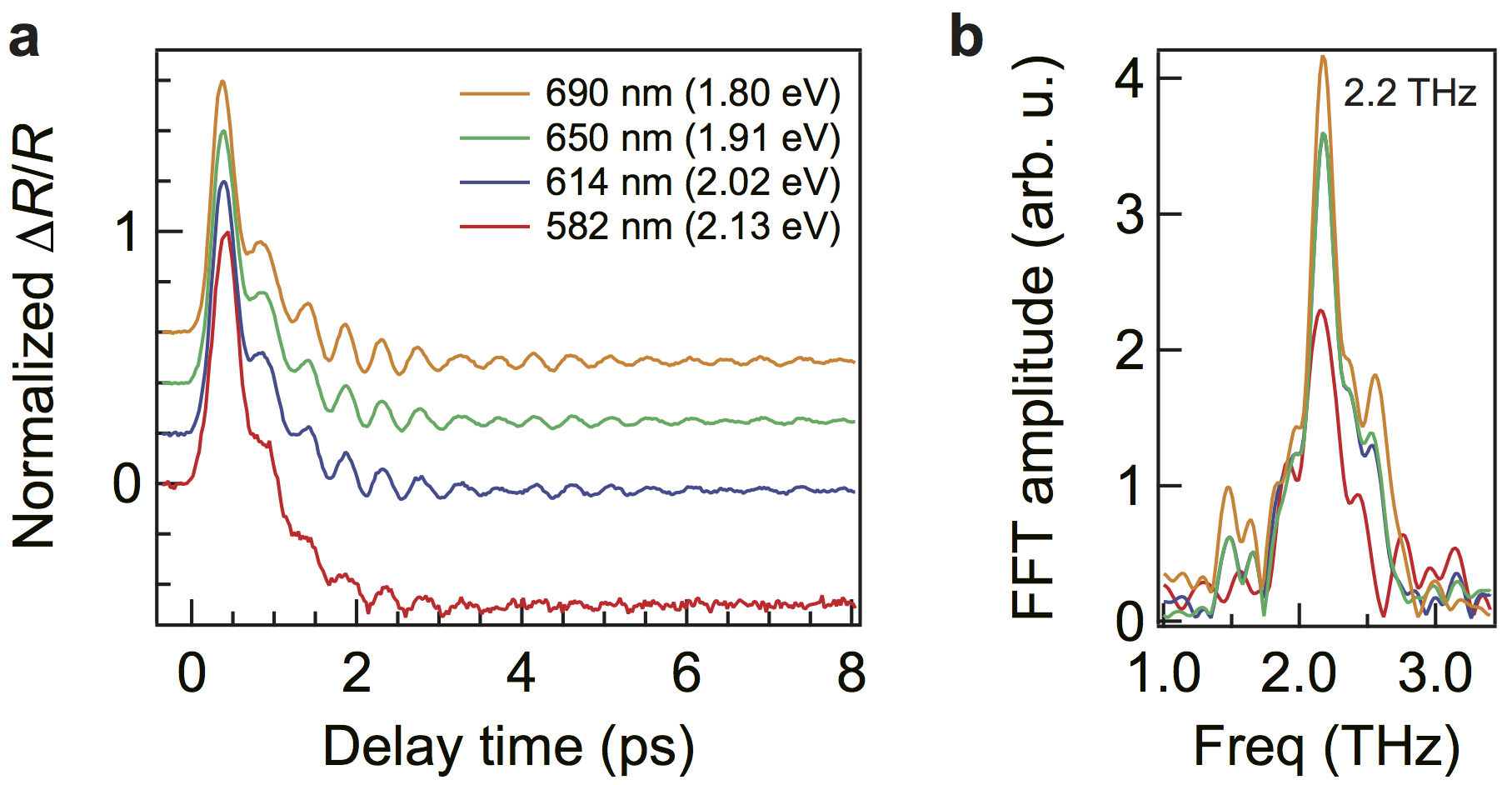}
    \caption{(Color online) Transient reflectivity with different probe photon wavelengths. (a)~Normalized $\Delta R/R$ traces at several probe photon wavelengths. The pump photons are the same (780\,nm, 1.59\,eV) with an identical excitation density of $F=1.2\times10^{20}$\,cm$^{-3}$. Traces are vertically offset for clarity. (b)~Fourier transform of the oscillatory component corresponding to traces in (a). The most prominent peak features the 2.2\,THz amplitude mode of the CDW. Traces are zero padded before the Fourier transform.}
    \label{fig:pp_wl}
\end{figure}

The transient reflectivity measurement employs several different probe photon energies in the white light continuum (500\,nm to 700\,nm). As different photon energies are sensitive to different inter-band transitions and scattering cross sections, we select the energy that is the most sensitive to the dynamics of the CDW amplitude. For this purpose, we compare the transient response at a fixed pump excitation density among several probe energies (Fig.\,\ref{fig:pp_wl}(a)). Traces are normalized to a maximum of unity for easy comparison. In Fig.\,\ref{fig:pp_wl}(b), we plot the Fourier transformed spectra of the coherent oscillatory component in Fig.\,\ref{fig:pp_wl}(a). In our measurement geometry of parallel polarizations between the pump and probe beams, the most prominent oscillation is the 2.2\,THz $A_{1g}$ CDW amplitude mode (AM). Among all probe energies, the 690\,nm (1.80\,eV) photon gives the highest AM peak (Fig.\,\ref{fig:pp_wl}(b)), making it the most suitable energy for probing the CDW gap dynamics in our available spectral range.

The observed photon energy dependence is expected if one examines the full optical conductivity spectrum, $\sigma_1(\omega)$, of LaTe$_3$, \cite{Sacchetti2006,Pfuner2009}. In the range of our white light continuum, we access the high frequency tail of the mid-infrared Lorentz harmonic oscillators in $\sigma_1(\omega)$, which compose the single particle peak of the CDW condensate. As the probe photon energy decreases, one moves closer to the center of the single particle peak. Therefore, the detection becomes more sensitive to a changing size of the CDW gap in the course of the photo-induced transition.

\subsection{VI. Time-dependent Landau theory}

A CDW transition is associated with both a modulation of electron density, described by the electronic order parameter $\psi_e$, and a periodic lattice distortion, described by the lattice order parameter $\psi_l$. If we are only interested in equilibrium properties, we can limit ourselves to considering either of these orders. However, as mobile electrons and heavy ions have very different dynamics, we need to consider the orders separately to capture the nonequilibrium state. Our subsequent analysis is based on the following Landau free-energy functional \cite{Goldenfeld1992,Schafer2010, Schaefer2014,Beaud2014}:
\begin{equation}
    {\cal W} = - \alpha_e |\psi_e|^2 + \frac{\beta}{2} |\psi_e|^4
- \zeta(\psi_e \psi_l^* + \psi_e^* \psi_l ) + \alpha_l |\psi_l|^2.
\end{equation}
This model implies that (i) non-linearity (the $|\psi_e|^4$ term) occurs due to electron-electron effects, (ii) the two orders are linearly coupled, and (iii) the lattice potential has a parabolic form originating from the elastic force. This potential can be straightforwardly generalized to non-homogeneous cases by including spatial derivatives, to cases where non-linearity is due to phononic effects, or to cases where more than one phonon mode is coupled to $\psi_e$. Despite its simplicity, this model contains all required physical ingredients and has relatively small number of free parameters. In addition, it has well described a number of recent experiments related to the CDW dynamics \cite{Yusupov2010,Schafer2010,Beaud2014}. Below, we also assume that photoexcitation acts as a homogeneous energy quench, allowing us to fix the order parameters to be real and to focus on the dynamical properties of the CDW amplitude.

The next step is to formulate the dynamical equations. The electron-electron interaction defines the fastest timescale ($\tau_e < 100$\,fs) in the system; on the phononic timescale, electrons instantly adjust themselves to the local value of the Landau potential. Hence, we impose overdamped dynamics on $\psi_e$, as done in the equilibrium time-dependent Landau formalism \cite{Goldenfeld1992}. On the other hand, the heavy ions behave more like classical objects. This allows us to introduce the following equations of motion:
\begin{equation}
    \frac{d \psi_e}{dt} \propto -\frac{\delta {\cal W}}{\psi_e}\text{ and }\frac{d^2 \psi_l}{dt^2} \propto -\frac{\delta {\cal W}}{\psi_l}.
\end{equation}
In normalized variables $x \equiv \psi_e(t) / \psi_e(T=0)$ and $y \equiv \psi_l(t) / \psi_l(T=0)$, these equations read
\begin{align}
&\tau_e \frac{d x}{dt} -\alpha(t) x + x^3 + \zeta_0 (x - y) = 0,\label{eq:TDGL1}\\
&\frac{1}{\omega_0^2} \frac{d^2 y}{dt^2} + (y - x) = 0. \label{eq:TDGL2}
\end{align}
Eqs.\,\eqref{eq:TDGL1}--\eqref{eq:TDGL2} imply that the effective force acting on the electronic (or lattice) order is $f_x \propto \alpha x - x^3 -\zeta_0 (x - y)$ (or $f_y \propto x - y$). $\zeta_0$ is the coupling strength between the electrons and the lattice; $\tau_e$ is the electron-electron scattering time that characterizes the relaxation within the electron subsystem; $\omega_0$ is the unrenormalized phonon frequency at $\mathbf{q}_\text{CDW}$; $\alpha(t)$ is the phenomenological Landau parameter that takes the value of $\alpha_0=(T_c - T_\text{env})/T_c$ at equilibrium, where $T_\text{env} = 300$\,K is the environment temperature. The time-dependence of $\alpha(t)$ is associated with the relaxation of excited quasiparticles after the laser pulse.

To model the photoexcitation event, we impose the following dynamics on the Landau potential,
\begin{equation}\label{eq:alpha}
\alpha(t) = \alpha_0 - \Theta(t)\kappa F e^{-t/\tau_0},
\end{equation}
where $\Theta(t)$ is the Heaviside step function, $F$ is the excitation density in the unit of photons/cm$^{3}$, and $\kappa$ is a constant of proportionality for normalizing $F$. $\tau_0$ is the intrinsic thermal relaxation time of the quasiparticles after photoexcitation. In the numerical calculation, $\tau_0 = 0.3$\,ps, taken from the transient reflectivity relaxation time at a small excitation density far below $T_c$ ~\cite{Zong2018,Yusupov2010,Yusupov2008}. Note that in Eq.\,\eqref{eq:alpha}, we neglect the temperature dependence of $\alpha(t)$ arising from transient lattice heating after photoexcitation. This is because based on heat capacity estimates, we expect a maximum lattice temperature rise of $\Delta T_{\rm max} \lesssim 100$\,K \cite{Zong2018}, which is small compared to the high transition temperature $T_c \approx 670$\,K \cite{Wang2014}.

To determine the values of $\tau_e$ and $\zeta_0$, we consider small fluctuations around the equilibrium state, $x(t) = x_\text{eq} + \delta x(t),\ y(t) = y_\text{eq} + \delta y(t)$, where $x_\text{eq} = y_\text{eq} =\sqrt{\alpha_0}$. For small $\delta x$ and $\delta y$, we can linearize Eqs.\,\eqref{eq:TDGL1}--\eqref{eq:TDGL2},
\begin{align}
&\tau_e \frac{d \delta x}{dt} + 2\alpha_0\delta x+ \zeta_0 (\delta x - \delta y) = 0,
\label{eq:TDGL_small1}\\
&\frac{1}{\omega_0^2} \frac{d^2 \delta y}{dt^2} + (\delta y - \delta x) = 0.\label{eq:TDGL_small2}
\end{align}
Using ansatz $\delta x = a e^{\lambda t}$ and $\delta y = b e^{\lambda t}$, we obtain a cubic equation for the eigenmode $\lambda$,
\begin{equation}\label{eq:freq}
\lambda^3 + \lambda^2 \frac{2 \alpha_0 + \zeta_0 }{\tau_e} + \lambda \omega_0^2 + \frac{2 \alpha_0 \omega_0^2}{\tau_e} = 0.
\end{equation}
The solution to Eq.\,\eqref{eq:freq} defines the CDW amplitude mode frequency $\omega_\text{AM}= \text{Im}\,\lambda$ and its damping $\gamma_\text{AM} = -\text{Re}\,\lambda$. Therefore, if one fixes the values of $\omega_\text{AM}$, $\gamma_\text{AM}$, and $\omega_0$, one can solve for $\tau_e$ and $\zeta_0$. Based on the AM measured in transient reflectivity (Fig.\,\ref{fig:pp_wl}(b)), the position and the width of the AM peak give $\omega_\text{AM} = 2.2\cdot(2\pi)$\,THz and $\gamma_\text{AM} = 0.2\cdot(2\pi)$\,THz. From previous Raman measurement, we expect $\omega_0=3.25\cdot(2\pi)$\,THz \cite{Eiter2013,Lavagnini2008}. These experimental parameters lead to an electron-electron scattering time $\tau_e=27$\,fs and electron-phonon coupling strength $\zeta_0=1.35$. It is worth noting that the numerically solved value of $\tau_e < 100$\,fs is expected for LaTe$_3$ -- a metallic system without strong electron-electron correlation effects \cite{Demsar1999,Prasankumar2016}.

\begin{figure}[th!]
    \includegraphics[scale=0.44]{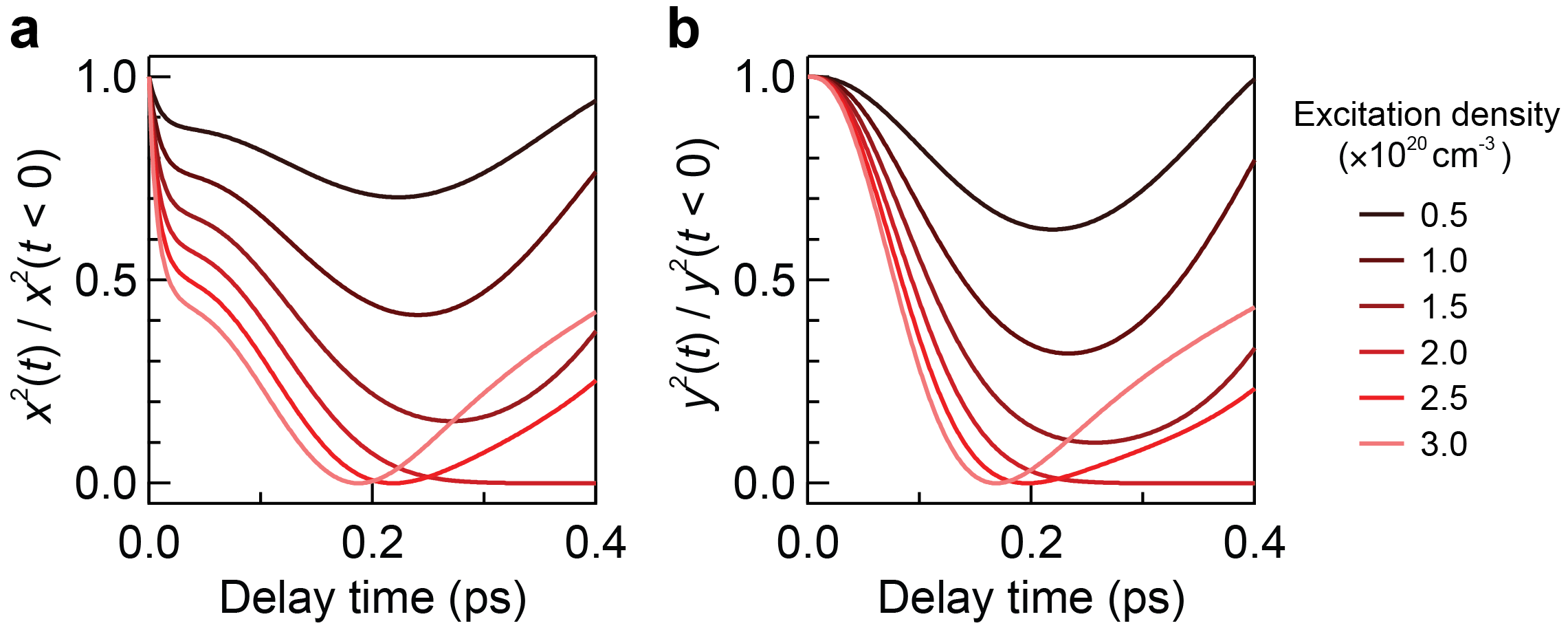}
    \caption{(Color online) Calculated dynamics of electronic (a) and lattice (b) parts of the order parameter, normalized to their respective values before photoexcitation ($t<0$).  Different curves denote different excitation densities in Eq.\,\eqref{eq:alpha}.}
    \label{fig:el_op}
\end{figure}

Using $x(t=0)=y(t=0)=\sqrt{\alpha_0}$ as the initial condition for Eqs.~(\ref{eq:TDGL1})--(\ref{eq:alpha}), we track the dynamics of the order parameter -- both electronic and lattice parts -- after photoexcitation for different excitation densities (Fig.\,\ref{fig:el_op}). For weak excitation, the order parameter ($x^2(t)$ or $y^2(t)$) is slightly suppressed and then quickly recovers to the initial value. Increasing the excitation density results in more pronounced suppression until the critical point is reached at $F_c$, where $x^2(t)$ and $y^2(t)$ asymptotically approach zero for the first time. It is interesting to observe that at even higher excitation density, the order parameters pass through zero in finite time. We associate this peculiarity with the fact that at $F_c$, the system is indeed in the critical regime where the Landau potential becomes flat (or equivalently, the effective forces $f_x \approx 0$ \emph{and} $f_y \approx 0$), and, as such, we expect an effect of dynamical slowing down -- see the discussion in the main text.

From Fig.\,\ref{fig:el_op}, the slowdown can be seen by tracking the position of the first minimum in $x^2(t)$, as plotted in Fig.\,\ref{fig:theory}(b). The result is similar if $y^2(t)$ is used instead of $x^2(t)$. We observe that indeed at $F_c$ the time needed to reach the minimum diverges. As discussed in the main text, this divergence is not stable against perturbations such as topological defects and is not manifested in the experiments. In the simulation, we adjusted $\kappa$ to match $F_c$ with experimentally determined $F_\text{melt}$.

It is worth mentioning that there is a kink in the dynamics of the electronic order in Fig.\,\ref{fig:el_op}(a) at $\sim30$\,fs, which is associated with the short electron-electron scattering time $\tau_e$. For the dynamics of the lattice order, no such kink is present in Fig.\,\ref{fig:el_op}(b) because the characteristic lattice response time is $\sim 2\pi/\omega_0 \approx 0.3$\,ps $\gg \tau_e$.

There are a few ingredients that can be added to the model to improve the quantitative agreement with the measurements. For example, $x(t)$ and $y(t)$ may vary spatially, mimicking photo-induced topological defects that smear the divergence in $\tau$ \cite{Goldenfeld1992,Yusupov2010,Zong2018}. In addition, the thermal relaxation of quasiparticles is at present represented by a fixed $\tau_0$, which is a simplification expected to work in a limited range of excitation density. An $F$-dependent $\tau_0$ may better represent the actual experiments. Lastly, only one phonon mode is considered in the current simulation, though other phonons have been shown to couple to the CDW order \cite{Yusupov2008}. Their incorporation may make the model more realistic.

Nonetheless, this minimal time-dependent Landau formalism demonstrates the qualitative features observed in our experiments, in particular, a slowdown in the order parameter dynamics in a regime far from equilibrium. With no adjustable parameters except $\kappa$ in our calculation, it is further encouraging to see a reasonable match in the absolute value of the CDW suppression time, $\tau$, between the calculation and the experiment. We also expect that this phenomenon of dynamical slowing down is general and, as such, will manifest in more complex models as a divergence in the time needed to suppress the order parameter.


\begin{thebibliography}{45}%
\makeatletter
\providecommand \@ifxundefined [1]{%
 \@ifx{#1\undefined}
}%
\providecommand \@ifnum [1]{%
 \ifnum #1\expandafter \@firstoftwo
 \else \expandafter \@secondoftwo
 \fi
}%
\providecommand \@ifx [1]{%
 \ifx #1\expandafter \@firstoftwo
 \else \expandafter \@secondoftwo
 \fi
}%
\providecommand \natexlab [1]{#1}%
\providecommand \enquote  [1]{``#1''}%
\providecommand \bibnamefont  [1]{#1}%
\providecommand \bibfnamefont [1]{#1}%
\providecommand \citenamefont [1]{#1}%
\providecommand \href@noop [0]{\@secondoftwo}%
\providecommand \href [0]{\begingroup \@sanitize@url \@href}%
\providecommand \@href[1]{\@@startlink{#1}\@@href}%
\providecommand \@@href[1]{\endgroup#1\@@endlink}%
\providecommand \@sanitize@url [0]{\catcode `\\12\catcode `\$12\catcode
  `\&12\catcode `\#12\catcode `\^12\catcode `\_12\catcode `\%12\relax}%
\providecommand \@@startlink[1]{}%
\providecommand \@@endlink[0]{}%
\providecommand \url  [0]{\begingroup\@sanitize@url \@url }%
\providecommand \@url [1]{\endgroup\@href {#1}{\urlprefix }}%
\providecommand \urlprefix  [0]{URL }%
\providecommand \Eprint [0]{\href }%
\providecommand \doibase [0]{http://dx.doi.org/}%
\providecommand \selectlanguage [0]{\@gobble}%
\providecommand \bibinfo  [0]{\@secondoftwo}%
\providecommand \bibfield  [0]{\@secondoftwo}%
\providecommand \translation [1]{[#1]}%
\providecommand \BibitemOpen [0]{}%
\providecommand \bibitemStop [0]{}%
\providecommand \bibitemNoStop [0]{.\EOS\space}%
\providecommand \EOS [0]{\spacefactor3000\relax}%
\providecommand \BibitemShut  [1]{\csname bibitem#1\endcsname}%
\let\auto@bib@innerbib\@empty
\bibitem [{\citenamefont {Collins}(1989)}]{Collins1989}%
  \BibitemOpen
  \bibfield  {author} {\bibinfo {author} {\bibfnamefont {M.~F.}\ \bibnamefont
  {Collins}},\ }\href@noop {} {\emph {\bibinfo {title} {{Magnetic Critical
  Scattering}}}}\ (\bibinfo  {publisher} {Oxford University Press},\ \bibinfo
  {year} {1989})\BibitemShut {NoStop}%
\bibitem [{\citenamefont {Goldenfeld}(1992)}]{Goldenfeld1992}%
  \BibitemOpen
  \bibfield  {author} {\bibinfo {author} {\bibfnamefont {N.}~\bibnamefont
  {Goldenfeld}},\ }\href@noop {} {\emph {\bibinfo {title} {Lectures on Phase
  Transitions and the Renormalization Group}}}\ (\bibinfo  {publisher}
  {Addison-Wesley},\ \bibinfo {year} {1992})\BibitemShut {NoStop}%
\bibitem [{\citenamefont {Horie}\ \emph {et~al.}(1987)\citenamefont {Horie},
  \citenamefont {Fukami},\ and\ \citenamefont {Mase}}]{Horie1987}%
  \BibitemOpen
  \bibfield  {author} {\bibinfo {author} {\bibfnamefont {Y.}~\bibnamefont
  {Horie}}, \bibinfo {author} {\bibfnamefont {T.}~\bibnamefont {Fukami}}, \
  and\ \bibinfo {author} {\bibfnamefont {S.}~\bibnamefont {Mase}},\ }\href
  {\doibase 10.1016/0038-1098(87)91100-8} {\bibfield  {journal} {\bibinfo
  {journal} {Solid State Commun.}\ }\textbf {\bibinfo {volume} {62}},\ \bibinfo
  {pages} {471} (\bibinfo {year} {1987})}\BibitemShut {NoStop}%
\bibitem [{\citenamefont {Zhu}\ \emph {et~al.}(2018)\citenamefont {Zhu} \emph
  {et~al.}}]{Zhu2018}%
  \BibitemOpen
  \bibfield  {author} {\bibinfo {author} {\bibfnamefont {Y.}~\bibnamefont
  {Zhu}} \emph {et~al.},\ }\href {\doibase 10.1038/s41467-018-04199-4}
  {\bibfield  {journal} {\bibinfo  {journal} {Nat. Commun.}\ }\textbf {\bibinfo
  {volume} {9}},\ \bibinfo {pages} {1799} (\bibinfo {year} {2018})}\BibitemShut
  {NoStop}%
\bibitem [{\citenamefont {Souletie}\ and\ \citenamefont
  {Tholence}(1985)}]{Souletie1985}%
  \BibitemOpen
  \bibfield  {author} {\bibinfo {author} {\bibfnamefont {J.}~\bibnamefont
  {Souletie}}\ and\ \bibinfo {author} {\bibfnamefont {J.~L.}\ \bibnamefont
  {Tholence}},\ }\href {\doibase 10.1103/PhysRevB.32.516} {\bibfield  {journal}
  {\bibinfo  {journal} {Phys. Rev. B}\ }\textbf {\bibinfo {volume} {32}},\
  \bibinfo {pages} {516} (\bibinfo {year} {1985})}\BibitemShut {NoStop}%
\bibitem [{\citenamefont {Lasjaunias}\ \emph {et~al.}(1994)\citenamefont
  {Lasjaunias}, \citenamefont {Biljakovi{\'{c}}}, \citenamefont {Nad'},
  \citenamefont {Monceau},\ and\ \citenamefont {Bechgaard}}]{Lasjaunias1994}%
  \BibitemOpen
  \bibfield  {author} {\bibinfo {author} {\bibfnamefont {J.~C.}\ \bibnamefont
  {Lasjaunias}}, \bibinfo {author} {\bibfnamefont {K.}~\bibnamefont
  {Biljakovi{\'{c}}}}, \bibinfo {author} {\bibfnamefont {F.}~\bibnamefont
  {Nad'}}, \bibinfo {author} {\bibfnamefont {P.}~\bibnamefont {Monceau}}, \
  and\ \bibinfo {author} {\bibfnamefont {K.}~\bibnamefont {Bechgaard}},\ }\href
  {\doibase 10.1103/PhysRevLett.72.1283} {\bibfield  {journal} {\bibinfo
  {journal} {Phys. Rev. Lett.}\ }\textbf {\bibinfo {volume} {72}},\ \bibinfo
  {pages} {1283} (\bibinfo {year} {1994})}\BibitemShut {NoStop}%
\bibitem [{\citenamefont {Strogatz}(2018)}]{strogatz2018}%
  \BibitemOpen
  \bibfield  {author} {\bibinfo {author} {\bibfnamefont {S.}~\bibnamefont
  {Strogatz}},\ }\href@noop {} {\emph {\bibinfo {title} {Nonlinear Dynamics and
  Chaos: With Applications to Physics, Biology, Chemistry, and Engineering}}}\
  (\bibinfo  {publisher} {CRC Press},\ \bibinfo {year} {2018})\BibitemShut
  {NoStop}%
\bibitem [{\citenamefont {Veraart}\ \emph {et~al.}(2012)\citenamefont {Veraart}
  \emph {et~al.}}]{Veraart2012}%
  \BibitemOpen
  \bibfield  {author} {\bibinfo {author} {\bibfnamefont {A.~J.}\ \bibnamefont
  {Veraart}} \emph {et~al.},\ }\href {\doibase 10.1038/nature10723} {\bibfield
  {journal} {\bibinfo  {journal} {Nature}\ }\textbf {\bibinfo {volume} {481}},\
  \bibinfo {pages} {357} (\bibinfo {year} {2012})}\BibitemShut {NoStop}%
\bibitem [{\citenamefont {Scheffer}\ \emph {et~al.}(2009)\citenamefont
  {Scheffer} \emph {et~al.}}]{Scheffer2009}%
  \BibitemOpen
  \bibfield  {author} {\bibinfo {author} {\bibfnamefont {M.}~\bibnamefont
  {Scheffer}} \emph {et~al.},\ }\href {\doibase 10.1038/nature08227} {\bibfield
   {journal} {\bibinfo  {journal} {Nature}\ }\textbf {\bibinfo {volume}
  {461}},\ \bibinfo {pages} {53} (\bibinfo {year} {2009})}\BibitemShut
  {NoStop}%
\bibitem [{\citenamefont {Collins}\ and\ \citenamefont
  {Teh}(1973)}]{Collins1973}%
  \BibitemOpen
  \bibfield  {author} {\bibinfo {author} {\bibfnamefont {M.~R.}\ \bibnamefont
  {Collins}}\ and\ \bibinfo {author} {\bibfnamefont {H.~C.}\ \bibnamefont
  {Teh}},\ }\href {\doibase 10.1103/PhysRevLett.30.781} {\bibfield  {journal}
  {\bibinfo  {journal} {Phys. Rev. Lett.}\ }\textbf {\bibinfo {volume} {30}},\
  \bibinfo {pages} {781} (\bibinfo {year} {1973})}\BibitemShut {NoStop}%
\bibitem [{\citenamefont {Iizumi}(1986)}]{Iizumi1986}%
  \BibitemOpen
  \bibfield  {author} {\bibinfo {author} {\bibfnamefont {M.}~\bibnamefont
  {Iizumi}},\ }\href {\doibase 10.1016/S0378-4363(86)80015-8} {\bibfield
  {journal} {\bibinfo  {journal} {Phys. B+C}\ }\textbf {\bibinfo {volume}
  {136}},\ \bibinfo {pages} {36} (\bibinfo {year} {1986})}\BibitemShut
  {NoStop}%
\bibitem [{\citenamefont {Cailleau}\ \emph {et~al.}(1979)\citenamefont
  {Cailleau}, \citenamefont {Heidemann},\ and\ \citenamefont
  {Zeyen}}]{Cailleau1979}%
  \BibitemOpen
  \bibfield  {author} {\bibinfo {author} {\bibfnamefont {H.}~\bibnamefont
  {Cailleau}}, \bibinfo {author} {\bibfnamefont {A.}~\bibnamefont {Heidemann}},
  \ and\ \bibinfo {author} {\bibfnamefont {C.~M.~E.}\ \bibnamefont {Zeyen}},\
  }\href {\doibase 10.1088/0022-3719/12/11/002} {\bibfield  {journal} {\bibinfo
   {journal} {J. Phys. C}\ }\textbf {\bibinfo {volume} {12}},\ \bibinfo {pages}
  {L411} (\bibinfo {year} {1979})}\BibitemShut {NoStop}%
\bibitem [{\citenamefont {Toudic}\ \emph {et~al.}(1986)\citenamefont {Toudic},
  \citenamefont {Cailleau}, \citenamefont {Lechner},\ and\ \citenamefont
  {Petry}}]{Toudic1986}%
  \BibitemOpen
  \bibfield  {author} {\bibinfo {author} {\bibfnamefont {B.}~\bibnamefont
  {Toudic}}, \bibinfo {author} {\bibfnamefont {H.}~\bibnamefont {Cailleau}},
  \bibinfo {author} {\bibfnamefont {R.~E.}\ \bibnamefont {Lechner}}, \ and\
  \bibinfo {author} {\bibfnamefont {W.}~\bibnamefont {Petry}},\ }\href
  {\doibase 10.1103/PhysRevLett.56.347} {\bibfield  {journal} {\bibinfo
  {journal} {Phys. Rev. Lett.}\ }\textbf {\bibinfo {volume} {56}},\ \bibinfo
  {pages} {347} (\bibinfo {year} {1986})}\BibitemShut {NoStop}%
\bibitem [{\citenamefont {Press}\ \emph {et~al.}(1974)\citenamefont {Press},
  \citenamefont {H{\"{u}}ller}, \citenamefont {Stiller}, \citenamefont
  {Stirling},\ and\ \citenamefont {Currat}}]{Press1974}%
  \BibitemOpen
  \bibfield  {author} {\bibinfo {author} {\bibfnamefont {W.}~\bibnamefont
  {Press}}, \bibinfo {author} {\bibfnamefont {A.}~\bibnamefont {H{\"{u}}ller}},
  \bibinfo {author} {\bibfnamefont {H.}~\bibnamefont {Stiller}}, \bibinfo
  {author} {\bibfnamefont {W.}~\bibnamefont {Stirling}}, \ and\ \bibinfo
  {author} {\bibfnamefont {R.}~\bibnamefont {Currat}},\ }\href {\doibase
  10.1103/PhysRevLett.32.1354} {\bibfield  {journal} {\bibinfo  {journal}
  {Phys. Rev. Lett.}\ }\textbf {\bibinfo {volume} {32}},\ \bibinfo {pages}
  {1354} (\bibinfo {year} {1974})}\BibitemShut {NoStop}%
\bibitem [{\citenamefont {Niermann}\ \emph {et~al.}(2015)\citenamefont
  {Niermann} \emph {et~al.}}]{Niermann2015}%
  \BibitemOpen
  \bibfield  {author} {\bibinfo {author} {\bibfnamefont {D.}~\bibnamefont
  {Niermann}} \emph {et~al.},\ }\href {\doibase 10.1103/PhysRevLett.114.037204}
  {\bibfield  {journal} {\bibinfo  {journal} {Phys. Rev. Lett.}\ }\textbf
  {\bibinfo {volume} {114}},\ \bibinfo {pages} {037204} (\bibinfo {year}
  {2015})}\BibitemShut {NoStop}%
\bibitem [{\citenamefont {Yusupov}\ \emph {et~al.}(2010)\citenamefont {Yusupov}
  \emph {et~al.}}]{Yusupov2010}%
  \BibitemOpen
  \bibfield  {author} {\bibinfo {author} {\bibfnamefont {R.}~\bibnamefont
  {Yusupov}} \emph {et~al.},\ }\href {https://doi.org/10.1038/nphys1738}
  {\bibfield  {journal} {\bibinfo  {journal} {Nat. Phys.}\ }\textbf {\bibinfo
  {volume} {6}},\ \bibinfo {pages} {681} (\bibinfo {year}
  {2010})}\BibitemShut {NoStop}%
\bibitem [{\citenamefont {Mertelj}\ \emph {et~al.}(2013)\citenamefont {Mertelj}
  \emph {et~al.}}]{Mertelj2013}%
  \BibitemOpen
  \bibfield  {author} {\bibinfo {author} {\bibfnamefont {T.}~\bibnamefont
  {Mertelj}} \emph {et~al.},\ }\href {\doibase 10.1103/PhysRevLett.110.156401}
  {\bibfield  {journal} {\bibinfo  {journal} {Phys. Rev. Lett.}\ }\textbf
  {\bibinfo {volume} {110}},\ \bibinfo {pages} {156401} (\bibinfo {year}
  {2013})}\BibitemShut {NoStop}%
\bibitem [{\citenamefont {Zong}\ \emph {et~al.}(2019)\citenamefont {Zong} \emph
  {et~al.}}]{Zong2018}%
  \BibitemOpen
  \bibfield  {author} {\bibinfo {author} {\bibfnamefont {A.}~\bibnamefont
  {Zong}} \emph {et~al.},\ }\href {https://doi.org/10.1038/s41567-018-0311-9}
  {\bibfield  {journal} {\bibinfo  {journal} {Nat. Phys.}\ }\textbf {\bibinfo
  {volume} {15}},\ \bibinfo {pages} {27} (\bibinfo {year} {2019})}\BibitemShut
  {NoStop}%
\bibitem [{\citenamefont {Zurek}(1996)}]{Zurek1996}%
  \BibitemOpen
  \bibfield  {author} {\bibinfo {author} {\bibfnamefont {W.}~\bibnamefont
  {Zurek}},\ }\href {\doibase 10.1016/S0370-1573(96)00009-9} {\bibfield
  {journal} {\bibinfo  {journal} {Phys. Rep.}\ }\textbf {\bibinfo {volume}
  {276}},\ \bibinfo {pages} {177} (\bibinfo {year} {1996})}\BibitemShut
  {NoStop}%
\bibitem [{\citenamefont {Chuang}\ \emph {et~al.}(1991)\citenamefont {Chuang},
  \citenamefont {Durrer}, \citenamefont {Turok},\ and\ \citenamefont
  {Yurke}}]{Chuang1991}%
  \BibitemOpen
  \bibfield  {author} {\bibinfo {author} {\bibfnamefont {I.}~\bibnamefont
  {Chuang}}, \bibinfo {author} {\bibfnamefont {R.}~\bibnamefont {Durrer}},
  \bibinfo {author} {\bibfnamefont {N.}~\bibnamefont {Turok}}, \ and\ \bibinfo
  {author} {\bibfnamefont {B.}~\bibnamefont {Yurke}},\ }\href {\doibase
  10.1126/science.251.4999.1336} {\bibfield  {journal} {\bibinfo  {journal}
  {Science}\ }\textbf {\bibinfo {volume} {251}},\ \bibinfo {pages} {1336}
  (\bibinfo {year} {1991})}\BibitemShut {NoStop}%
\bibitem [{\citenamefont {Bowick}\ \emph {et~al.}(1994)\citenamefont {Bowick},
  \citenamefont {Chandar}, \citenamefont {Schiff},\ and\ \citenamefont
  {Srivastava}}]{Bowick1994}%
  \BibitemOpen
  \bibfield  {author} {\bibinfo {author} {\bibfnamefont {M.}~\bibnamefont
  {Bowick}}, \bibinfo {author} {\bibfnamefont {L.}~\bibnamefont {Chandar}},
  \bibinfo {author} {\bibfnamefont {E.}~\bibnamefont {Schiff}}, \ and\ \bibinfo
  {author} {\bibfnamefont {A.}~\bibnamefont {Srivastava}},\ }\href {\doibase
  10.1126/science.263.5149.943} {\bibfield  {journal} {\bibinfo  {journal}
  {Science}\ }\textbf {\bibinfo {volume} {263}},\ \bibinfo {pages} {943}
  (\bibinfo {year} {1994})}\BibitemShut {NoStop}%
\bibitem [{\citenamefont {Tomeljak}\ \emph {et~al.}(2009)\citenamefont
  {Tomeljak} \emph {et~al.}}]{Demsar2009}%
  \BibitemOpen
  \bibfield  {author} {\bibinfo {author} {\bibfnamefont {A.}~\bibnamefont
  {Tomeljak}} \emph {et~al.},\ }\href {\doibase 10.1103/PhysRevLett.102.066404}
  {\bibfield  {journal} {\bibinfo  {journal} {Phys. Rev. Lett.}\ }\textbf
  {\bibinfo {volume} {102}},\ \bibinfo {pages} {066404} (\bibinfo {year}
  {2009})}\BibitemShut {NoStop}%
\bibitem [{\citenamefont {Demsar}\ \emph {et~al.}(1999)\citenamefont {Demsar},
  \citenamefont {Biljakovi\ifmmode~\acute{c}\else \'{c}\fi{}},\ and\
  \citenamefont {Mihailovic}}]{Demsar1999}%
  \BibitemOpen
  \bibfield  {author} {\bibinfo {author} {\bibfnamefont {J.}~\bibnamefont
  {Demsar}}, \bibinfo {author} {\bibfnamefont {K.}~\bibnamefont
  {Biljakovi\ifmmode~\acute{c}\else \'{c}\fi{}}}, \ and\ \bibinfo {author}
  {\bibfnamefont {D.}~\bibnamefont {Mihailovic}},\ }\href {\doibase
  10.1103/PhysRevLett.83.800} {\bibfield  {journal} {\bibinfo  {journal} {Phys.
  Rev. Lett.}\ }\textbf {\bibinfo {volume} {83}},\ \bibinfo {pages} {800}
  (\bibinfo {year} {1999})}\BibitemShut {NoStop}%
\bibitem [{\citenamefont {Kabanov}\ \emph {et~al.}(1999)\citenamefont
  {Kabanov}, \citenamefont {Demsar}, \citenamefont {Podobnik},\ and\
  \citenamefont {Mihailovic}}]{Kabanov1999}%
  \BibitemOpen
  \bibfield  {author} {\bibinfo {author} {\bibfnamefont {V.}~\bibnamefont
  {Kabanov}}, \bibinfo {author} {\bibfnamefont {J.}~\bibnamefont {Demsar}},
  \bibinfo {author} {\bibfnamefont {B.}~\bibnamefont {Podobnik}}, \ and\
  \bibinfo {author} {\bibfnamefont {D.}~\bibnamefont {Mihailovic}},\ }\href
  {\doibase 10.1103/PhysRevB.59.1497} {\bibfield  {journal} {\bibinfo
  {journal} {Phys. Rev. B}\ }\textbf {\bibinfo {volume} {59}},\ \bibinfo
  {pages} {1497} (\bibinfo {year} {1999})}\BibitemShut {NoStop}%
\bibitem [{\citenamefont {Yusupov}\ \emph {et~al.}(2008)\citenamefont
  {Yusupov}, \citenamefont {Mertelj}, \citenamefont {Chu}, \citenamefont
  {Fisher},\ and\ \citenamefont {Mihailovic}}]{Yusupov2008}%
  \BibitemOpen
  \bibfield  {author} {\bibinfo {author} {\bibfnamefont {R.~V.}\ \bibnamefont
  {Yusupov}}, \bibinfo {author} {\bibfnamefont {T.}~\bibnamefont {Mertelj}},
  \bibinfo {author} {\bibfnamefont {J.-H.}\ \bibnamefont {Chu}}, \bibinfo
  {author} {\bibfnamefont {I.~R.}\ \bibnamefont {Fisher}}, \ and\ \bibinfo
  {author} {\bibfnamefont {D.}~\bibnamefont {Mihailovic}},\ }\href {\doibase
  10.1103/PhysRevLett.101.246402} {\bibfield  {journal} {\bibinfo  {journal}
  {Phys. Rev. Lett.}\ }\textbf {\bibinfo {volume} {101}},\ \bibinfo {pages}
  {246402} (\bibinfo {year} {2008})}\BibitemShut {NoStop}%
\bibitem [{\citenamefont {Ru}\ \emph {et~al.}(2008)\citenamefont {Ru} \emph
  {et~al.}}]{Ru2008}%
  \BibitemOpen
  \bibfield  {author} {\bibinfo {author} {\bibfnamefont {N.}~\bibnamefont {Ru}}
  \emph {et~al.},\ }\href {\doibase 10.1103/PhysRevB.77.035114} {\bibfield
  {journal} {\bibinfo  {journal} {Phys. Rev. B}\ }\textbf {\bibinfo {volume}
  {77}},\ \bibinfo {pages} {035114} (\bibinfo {year} {2008})}\BibitemShut
  {NoStop}%
\bibitem [{\citenamefont {Hu}\ \emph {et~al.}(2014)\citenamefont {Hu},
  \citenamefont {Cheng}, \citenamefont {Yuan}, \citenamefont {Dong},\ and\
  \citenamefont {Wang}}]{Wang2014}%
  \BibitemOpen
  \bibfield  {author} {\bibinfo {author} {\bibfnamefont {B.}~\bibnamefont
  {Hu}}, \bibinfo {author} {\bibfnamefont {B.}~\bibnamefont {Cheng}}, \bibinfo
  {author} {\bibfnamefont {R.}~\bibnamefont {Yuan}}, \bibinfo {author}
  {\bibfnamefont {T.}~\bibnamefont {Dong}}, \ and\ \bibinfo {author}
  {\bibfnamefont {N.}~\bibnamefont {Wang}},\ }\href {\doibase
  10.1103/PhysRevB.90.085105} {\bibfield  {journal} {\bibinfo  {journal} {Phys.
  Rev. B}\ }\textbf {\bibinfo {volume} {90}},\ \bibinfo {pages} {085105}
  (\bibinfo {year} {2014})}\BibitemShut {NoStop}%
\bibitem [{\citenamefont {Brouet}\ \emph {et~al.}(2008)\citenamefont {Brouet}
  \emph {et~al.}}]{Brouet2008}%
  \BibitemOpen
  \bibfield  {author} {\bibinfo {author} {\bibfnamefont {V.}~\bibnamefont
  {Brouet}} \emph {et~al.},\ }\href {\doibase 10.1103/PhysRevB.77.235104}
  {\bibfield  {journal} {\bibinfo  {journal} {Phys. Rev. B}\ }\textbf {\bibinfo
  {volume} {77}},\ \bibinfo {pages} {235104} (\bibinfo {year}
  {2008})}\BibitemShut {NoStop}%
\bibitem [{\citenamefont {Moore}\ \emph {et~al.}(2016)\citenamefont {Moore}
  \emph {et~al.}}]{Moore2016}%
  \BibitemOpen
  \bibfield  {author} {\bibinfo {author} {\bibfnamefont {R.}~\bibnamefont
  {Moore}} \emph {et~al.},\ }\href {\doibase 10.1103/PhysRevB.93.024304}
  {\bibfield  {journal} {\bibinfo  {journal} {Phys. Rev. B}\ }\textbf {\bibinfo
  {volume} {93}},\ \bibinfo {pages} {024304} (\bibinfo {year}
  {2016})}\BibitemShut {NoStop}%
\bibitem [{\citenamefont {Schmitt}\ \emph {et~al.}(2008)\citenamefont {Schmitt}
  \emph {et~al.}}]{Schmitt2008}%
  \BibitemOpen
  \bibfield  {author} {\bibinfo {author} {\bibfnamefont {F.}~\bibnamefont
  {Schmitt}} \emph {et~al.},\ }\href {\doibase 10.1126/science.1160778}
  {\bibfield  {journal} {\bibinfo  {journal} {Science}\ }\textbf {\bibinfo
  {volume} {321}},\ \bibinfo {pages} {1649} (\bibinfo {year}
  {2008})}\BibitemShut {NoStop}%
\bibitem [{SM()}]{SM}%
  \BibitemOpen
  \href@noop {} {}\bibinfo {note} {See supplemental materials for more
  details.}\BibitemShut {Stop}%
\bibitem [{\citenamefont {Hellmann}\ \emph {et~al.}(2012)\citenamefont
  {Hellmann} \emph {et~al.}}]{Hellmann2012}%
  \BibitemOpen
  \bibfield  {author} {\bibinfo {author} {\bibfnamefont {S.}~\bibnamefont
  {Hellmann}} \emph {et~al.},\ }\href
  {https://www.nature.com/articles/ncomms2078} {\bibfield  {journal} {\bibinfo
  {journal} {Nat. Commun.}\ }\textbf {\bibinfo {volume} {3}},\ \bibinfo {pages}
  {1069} (\bibinfo {year} {2012})}\BibitemShut {NoStop}%
\bibitem [{\citenamefont {Trigo}\ \emph {et~al.}(2018)\citenamefont {Trigo}
  \emph {et~al.}}]{Trigo2018}%
  \BibitemOpen
  \bibfield  {author} {\bibinfo {author} {\bibfnamefont {M.}~\bibnamefont
  {Trigo}} \emph {et~al.},\ }\href {http://arxiv.org/abs/1809.09799} {\bibfield
   {journal} {\bibinfo  {journal} {arXiv:1809.09799}\ } (\bibinfo {year}
  {2018})}\BibitemShut {NoStop}%
\bibitem [{\citenamefont {Demsar}\ and\ \citenamefont
  {Dekorsy}(2016)}]{Prasankumar2016}%
  \BibitemOpen
  \bibfield  {author} {\bibinfo {author} {\bibfnamefont {J.}~\bibnamefont
  {Demsar}}\ and\ \bibinfo {author} {\bibfnamefont {T.}~\bibnamefont
  {Dekorsy}},\ }in\ \href
  {https://www.crcpress.com/Optical-Techniques-for-Solid-State-Materials-Characterization/Prasankumar-Taylor/p/book/9781439815373}
  {\emph {\bibinfo {booktitle} {Optical Techniques for Solid-State Materials
  Characterization}}},\ \bibinfo {editor} {edited by\ \bibinfo {editor}
  {\bibfnamefont {R.}~\bibnamefont {Prasankumar}}\ and\ \bibinfo {editor}
  {\bibfnamefont {A.}~\bibnamefont {Taylor}}}\ (\bibinfo  {publisher} {CRC
  Press},\ \bibinfo {year} {2016})\ pp.\ \bibinfo {pages}
  {291--328}\BibitemShut {NoStop}%
\bibitem [{\citenamefont {Weathersby}\ \emph {et~al.}(2015)\citenamefont
  {Weathersby} \emph {et~al.}}]{Weathersby2015}%
  \BibitemOpen
  \bibfield  {author} {\bibinfo {author} {\bibfnamefont {S.~P.}\ \bibnamefont
  {Weathersby}} \emph {et~al.},\ }\href {\doibase 10.1063/1.4926994} {\bibfield
   {journal} {\bibinfo  {journal} {Rev. Sci. Instrum.}\ }\textbf {\bibinfo
  {volume} {86}},\ \bibinfo {pages} {073702} (\bibinfo {year}
  {2015})}\BibitemShut {NoStop}%
\bibitem [{\citenamefont {Shen}\ \emph {et~al.}(2018)\citenamefont {Shen} \emph
  {et~al.}}]{Shen2018}%
  \BibitemOpen
  \bibfield  {author} {\bibinfo {author} {\bibfnamefont {X.}~\bibnamefont
  {Shen}} \emph {et~al.},\ }\href {\doibase 10.1016/j.ultramic.2017.08.019}
  {\bibfield  {journal} {\bibinfo  {journal} {Ultramicroscopy}\ }\textbf
  {\bibinfo {volume} {184}},\ \bibinfo {pages} {172} (\bibinfo {year}
  {2018})}\BibitemShut {NoStop}%
\bibitem [{\citenamefont {Sacchetti}\ \emph {et~al.}(2006)\citenamefont
  {Sacchetti}, \citenamefont {Degiorgi}, \citenamefont {Giamarchi},
  \citenamefont {Ru},\ and\ \citenamefont {Fisher}}]{Sacchetti2006}%
  \BibitemOpen
  \bibfield  {author} {\bibinfo {author} {\bibfnamefont {A.}~\bibnamefont
  {Sacchetti}}, \bibinfo {author} {\bibfnamefont {L.}~\bibnamefont {Degiorgi}},
  \bibinfo {author} {\bibfnamefont {T.}~\bibnamefont {Giamarchi}}, \bibinfo
  {author} {\bibfnamefont {N.}~\bibnamefont {Ru}}, \ and\ \bibinfo {author}
  {\bibfnamefont {I.}~\bibnamefont {Fisher}},\ }\href {\doibase
  10.1103/PhysRevB.74.125115} {\bibfield  {journal} {\bibinfo  {journal} {Phys.
  Rev. B}\ }\textbf {\bibinfo {volume} {74}},\ \bibinfo {pages} {125115}
  (\bibinfo {year} {2006})}\BibitemShut {NoStop}%
\bibitem [{\citenamefont {Pfuner}\ \emph {et~al.}(2009)\citenamefont {Pfuner}
  \emph {et~al.}}]{Pfuner2009}%
  \BibitemOpen
  \bibfield  {author} {\bibinfo {author} {\bibfnamefont {F.}~\bibnamefont
  {Pfuner}} \emph {et~al.},\ }\href {\doibase 10.1016/j.physb.2008.11.052}
  {\bibfield  {journal} {\bibinfo  {journal} {Phys. B}\ }\textbf {\bibinfo
  {volume} {404}},\ \bibinfo {pages} {533} (\bibinfo {year}
  {2009})}\BibitemShut {NoStop}%
\bibitem [{\citenamefont {Li}\ \emph {et~al.}(2018)\citenamefont {Li} \emph
  {et~al.}}]{RKLi2018}%
  \BibitemOpen
  \bibfield  {author} {\bibinfo {author} {\bibfnamefont {R.~K.}\ \bibnamefont
  {Li}} \emph {et~al.},\ }\href {https://arxiv.org/abs/1805.01979} {\bibfield
  {journal} {\bibinfo  {journal} {arXiv:1805.01979}\ } (\bibinfo {year}
  {2018})}\BibitemShut {NoStop}%
\bibitem [{\citenamefont {Ofori-Okai}\ \emph {et~al.}(2018)\citenamefont
  {Ofori-Okai} \emph {et~al.}}]{Ofori-Okai2018}%
  \BibitemOpen
  \bibfield  {author} {\bibinfo {author} {\bibfnamefont {B.}~\bibnamefont
  {Ofori-Okai}} \emph {et~al.},\ }\href {\doibase
  10.1088/1748-0221/13/06/P06014} {\bibfield  {journal} {\bibinfo  {journal}
  {J. Instrum.}\ }\textbf {\bibinfo {volume} {13}},\ \bibinfo {pages} {P06014}
  (\bibinfo {year} {2018})}\BibitemShut {NoStop}%
\bibitem [{\citenamefont {Sch{\"{a}}fer}\ \emph {et~al.}(2010)\citenamefont
  {Sch{\"{a}}fer}, \citenamefont {Kabanov}, \citenamefont {Beyer},
  \citenamefont {Biljakovic},\ and\ \citenamefont {Demsar}}]{Schafer2010}%
  \BibitemOpen
  \bibfield  {author} {\bibinfo {author} {\bibfnamefont {H.}~\bibnamefont
  {Sch{\"{a}}fer}}, \bibinfo {author} {\bibfnamefont {V.~V.}\ \bibnamefont
  {Kabanov}}, \bibinfo {author} {\bibfnamefont {M.}~\bibnamefont {Beyer}},
  \bibinfo {author} {\bibfnamefont {K.}~\bibnamefont {Biljakovic}}, \ and\
  \bibinfo {author} {\bibfnamefont {J.}~\bibnamefont {Demsar}},\ }\href
  {\doibase 10.1103/PhysRevLett.105.066402} {\bibfield  {journal} {\bibinfo
  {journal} {Phys. Rev. Lett.}\ }\textbf {\bibinfo {volume} {105}},\ \bibinfo
  {pages} {066402} (\bibinfo {year} {2010})}\BibitemShut {NoStop}%
\bibitem [{\citenamefont {Schaefer}\ \emph {et~al.}(2014)\citenamefont
  {Schaefer}, \citenamefont {Kabanov},\ and\ \citenamefont
  {Demsar}}]{Schaefer2014}%
  \BibitemOpen
  \bibfield  {author} {\bibinfo {author} {\bibfnamefont {H.}~\bibnamefont
  {Schaefer}}, \bibinfo {author} {\bibfnamefont {V.~V.}\ \bibnamefont
  {Kabanov}}, \ and\ \bibinfo {author} {\bibfnamefont {J.}~\bibnamefont
  {Demsar}},\ }\href {\doibase 10.1103/PhysRevB.89.045106} {\bibfield
  {journal} {\bibinfo  {journal} {Phys. Rev. B}\ }\textbf {\bibinfo {volume}
  {89}},\ \bibinfo {pages} {045106} (\bibinfo {year} {2014})}\BibitemShut
  {NoStop}%
\bibitem [{\citenamefont {Beaud}\ \emph {et~al.}(2014)\citenamefont {Beaud}
  \emph {et~al.}}]{Beaud2014}%
  \BibitemOpen
  \bibfield  {author} {\bibinfo {author} {\bibfnamefont {P.}~\bibnamefont
  {Beaud}} \emph {et~al.},\ }\href {\doibase 10.1038/nmat4046} {\bibfield
  {journal} {\bibinfo  {journal} {Nat. Mater.}\ }\textbf {\bibinfo {volume}
  {13}},\ \bibinfo {pages} {923} (\bibinfo {year} {2014})}\BibitemShut
  {NoStop}%
\bibitem [{\citenamefont {Eiter}\ \emph {et~al.}(2013)\citenamefont {Eiter}
  \emph {et~al.}}]{Eiter2013}%
  \BibitemOpen
  \bibfield  {author} {\bibinfo {author} {\bibfnamefont {H.-M.}\ \bibnamefont
  {Eiter}} \emph {et~al.},\ }\href {\doibase 10.1073/pnas.1214745110}
  {\bibfield  {journal} {\bibinfo  {journal} {Proc. Natl. Acad. Sci.}\ }\textbf
  {\bibinfo {volume} {110}},\ \bibinfo {pages} {64} (\bibinfo {year}
  {2013})}\BibitemShut {NoStop}%
\bibitem [{\citenamefont {Lavagnini}\ \emph {et~al.}(2008)\citenamefont
  {Lavagnini} \emph {et~al.}}]{Lavagnini2008}%
  \BibitemOpen
  \bibfield  {author} {\bibinfo {author} {\bibfnamefont {M.}~\bibnamefont
  {Lavagnini}} \emph {et~al.},\ }\href {\doibase 10.1103/PhysRevB.78.201101}
  {\bibfield  {journal} {\bibinfo  {journal} {Phys. Rev. B}\ }\textbf {\bibinfo
  {volume} {78}},\ \bibinfo {pages} {201101} (\bibinfo {year}
  {2008})}\BibitemShut {NoStop}%
\end{thebibliography}
\end{document}